\pdfoutput=1
\documentclass[12pt]{iopart}

\usepackage{graphicx}
\usepackage{dcolumn}
\usepackage{bm}
\usepackage{dsfont}
\usepackage{iopams}

\newcommand{\boldphi}{{\mbox{\boldmath $\varphi$}}}

\newcommand{\bC}{{\bf C}}
\newcommand{\cH}{{\cal H}}
\newcommand{\bL}{{\bf L}}

\newcommand{\bM}{{\bf M}}
\newcommand{\bN}{{\bf N}}
\newcommand{\boldPhi}{{\mbox{\boldmath $\Phi$}}}
\newcommand{\bS}{{\bf S}}
\newcommand{\bI}{{\bf I}}
\newcommand{\bF}{{\bf F}}
\newcommand{\bQ}{{\bf Q}}

\newcommand{\ba}{\begin{array}}
\newcommand{\ea}{\end{array}}

\begin{document}

\title{Efficient one- and two-qubit pulsed gates for an oscillator stabilized Josephson qubit}
\author{Frederico Brito, David P. DiVincenzo, Roger H. Koch\footnote{Deceased}, and Matthias Steffen}

\address{ IBM T. J. Watson Research Center, P. O. Box 218, Yorktown Heights, NY 10598 USA
}%

\date{\today}

\begin{abstract}
We present theoretical schemes for performing high-fidelity one- and two-qubit
pulsed gates for a superconducting flux qubit. The ``IBM qubit"
consists of three Josephson junctions, three loops, and a
superconducting transmission line. Assuming a fixed inductive
qubit-qubit coupling, we show that the effective qubit-qubit
interaction is tunable by changing the applied fluxes, and can be
made negligible, allowing one to perform high fidelity single
qubit gates.  Our schemes are tailored to alleviate errors due to
1/f noise; we find gates with only 1\% loss of fidelity due to
this source, for pulse times in the range of 20-30ns for one-qubit
gates (Z rotations, Hadamard), and 60ns for a two-qubit gate
(controlled-Z). Our relaxation and dephasing time estimates
indicate a comparable loss of fidelity from this source. The
control of leakage plays an important role in the design of our
shaped pulses, preventing shorter pulse times. However, we
have found that imprecision in the control of the quantum phase
plays the major role in the limitation of the fidelity of our
gates.
\end{abstract}

\maketitle

\section{\label{introduction}Introduction}

Superconducting circuits containing Josephson
junctions \cite{Makhlin:2001aa,tinkham} are widely recognized to
be promising systems for the physical implementation of quantum
bits. These systems can be made and operated using
well-established experimental techniques, and they have the clear
potential, in principle, for scalability. Important experimental
milestones in coupled superconducting qubits, including the
observation of two-qubit gates \cite{Yamamoto:2003aa,
Plantenberg:2007aa} and the measurement of entanglement
\cite{Steffen:2006aa}, have already been reached. But because
superconducting qubits are condensed matter systems, it has been a
hard task to isolate them from their environment. Because of that,
these systems tend to suffer from short coherence times, which has
imposed serious limitations on achieving very high fidelity gates.

In this paper, we report a theoretical study of a universal set of
one- and two-qubit gates implemented using only shaped dc flux
pulses for an oscillator stabilized flux qubit.  We introduce a
simplified but accurate model to describe the dynamics of the
lowest states of the qubit as a function of the external control
parameters. This model provides a simpler way to analyze the
physics of the problem, helping to make it easy to see what
operations are necessary for performing the desired quantum gates.
As we shall see, there is a trade-off between the speed of a gate
and the amount of leakage produced by it (leakage = evolution of
states out of the 0-1 computational basis).  Smart choices for the
shape of the pulses are required to keep leakage at a tolerable
level, such that the gate fidelities are not compromised by this process \cite{plenio,Fazio}.  We have been able to keep leakage at the 0.12\% level for
gate times of order of few tens of ns.

The other important goal of the present work is to consider the
effect of low-frequency noise during the gate operation and in the
memory state. When designing quantum gates, we search for the best
pulse paths such that the loss of fidelity due to the low-frequency
noise is minimized.  We find that for all gates of interest, the
loss of fidelity due to low frequency noise is never greater than
1\%.  Achieving this low level of infidelity requires a careful
use of symmetries (so that errors accumulated in the first half of
a pulse can be cancelled out in the second, for example) and of
``sweet spots'' \cite{vion,wallraff} (points in the control parameter space that are
first-order insensitive to fluctuations).

The last part of this paper analyzes a pulsed two-qubit gate, the
controlled phase gate. An interesting feature of the coupling used
in this gate scheme is that, although the physical inductive
coupling between the qubits is assumed fixed, the effective
qubit-qubit interaction is tunable as a function of the control
parameters on both qubits, due to the change of character of the bare
qubit states as the control parameters are varied.  In fact, the
effective qubit-qubit interaction can be made negligible in large
regions of the flux space, allowing us to perform high fidelity
one-qubit gates for the two-qubit system without extremely
stringent control over electrical cross-talk.

The outline of this paper is as follows. In sections
\ref{systemhamiltonian} and \ref{4level} we discuss the physics of our system, and
present the simplified model used to simulate the dynamics of the
lowest lying levels of an oscillator stabilized flux qubit. The
Appendix carries out a detailed derivation of the system
Hamiltonian, and the regime of validity of the simplified model is
discussed. Section \ref{gates} presents the different schemes
designed to perform the following basic single-qubit operations:
measurement in the standard basis (0/1), measurement in the
conjugate basis (+/-), Z-rotation gates, and the Hadamard gate. The
fidelity of all these gates is given as a function of unwanted
shifts from the optimal point of operation in flux and time
synchronization of the pulses. In addition, we characterize
the noise through the operator-sum representation of the system superoperator. In
section \ref{twoqubit}, the two-qubit system is analyzed. The form
of and the reasons for an effective tunable interaction are
discussed. Then, a gate in the equivalence class of the
controlled-Z gate is proposed, and its fidelity and a characterization of the nature of the noise are presented. Finally, section \ref{conclusion}
gives some conclusions.

\section{\label{systemhamiltonian}System Hamiltonian}

\begin{figure}[t!]
\begin{center}\includegraphics[ width=0.9\columnwidth,
 keepaspectratio]{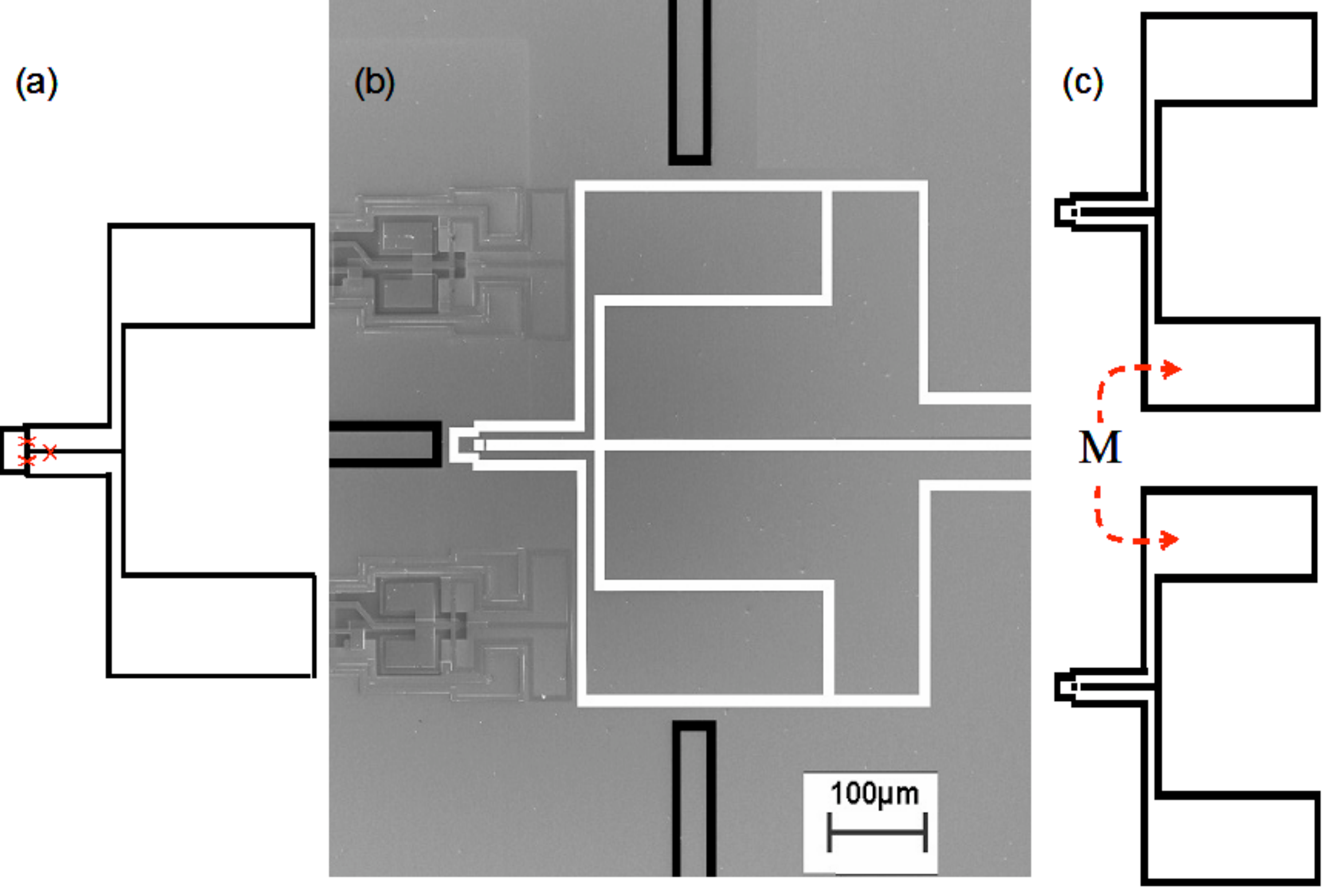}\end{center}
    \caption{(a) Schematic layout of the {\it bare} IBM qubit. This device consists of three loops and three Josephson                  junctions (indicated by red crosses), and operates in the flux regime. (b) Picture of the qubit. The white tracks                   highlight the bare qubit (a),
    and the transmission line coupled to it (white tracks leaving the
    right-hand side of the picture). In order to operate the qubit, flux lines
    (highlighted in black tracks) are used to change the total fluxes
    threading the loops. Readout SQUIDs (two such structures are
    shown at
    the left-hand side of the picture) perform the measurement of the state of
    the qubit. (c) Proposed scheme of the two qubit system. The
    qubit-qubit interaction is assumed to arise via the indicated mutual
    inductance between the two big loops\cite{PhysRevB.60.15398}. This results in
    a qubit-qubit interaction Hamiltonian of the form
    $\hat{\sigma}_z\otimes\hat{\sigma}_z$. Physical parameters for this qubit: the capacitances and critical currents of the junctions are assumed $C=$10fF and $I_c=1.3\mu$A; $L_T=5.6$nH, $L_1=32$pH and $L_3=680$pH  are the transmission line, the small and big loop inductances, respectively. The mutual inductance between qubit-transmission line, small loop-control flux line and big loop-bias flux line are respectively 200pH, 0.8pH and 0.5pH. The transmission line is designed to have a fundamental mode frequency of $\omega_T=2\pi\times3.1$GHz. Finally, the qubit-qubit mutual inductance is $M=$12pH.}
    \label{qubit}
\end{figure}

Our analysis will be focused on the IBM qubit
\cite{Koch:2005aa,Koch:2006aa}. This device, shown in
Fig.\ref{qubit} consists of a {\em bare qubit}, which is a type of
flux qubit containing three Josephson junctions and three loops,
and a high-quality superconducting {\em transmission line}
\cite{blais,Koch:2006aa,wallraff}. The bare qubit is subject to
external control via flux lines which change the total magnetic
fluxes threading the loops. As previously reported
\cite{DiVincenzo:2006aa}, and summarized in the Appendix, the bare
qubit has a gradiometric structure, and hence its behavior, to
good approximation (see Appendix for the validity of this
assumption), is only a function of the difference of the magnetic
flux in the two large loops, which we denote as
$\epsilon\equiv\Phi-\Phi_{\rm p}$. Whenever $\epsilon$ is an
integer multiple of the flux quantum $\Phi_0=h/2e$, the system
potential has a perfectly symmetric structure. These lines in the
flux space are referred as the ``S Lines'' (S for symmetric).

The bare qubit has two control parameters that define its flux
space: one, $\epsilon$, causes departures from the parity symmetry
just described, and the other, the so called {\em control flux}
$\tilde{\Phi}_c$, changes the structure of the qubit potential,
varying the height of the potential barrier between the two
classically stable states.  As $\tilde{\Phi}_c$ is changed, the
system potential passes from a double-well to a single-well
structure. This change in the form of the potential as a function
of a control parameter produces an essential feature in the
quantum behavior of the system: a well-defined change in the
character of the qubit eigenstates. While for the double-well
regime the qubit eigenstates are almost localized orbitals in the
left and right wells, as the single well regime is approached the
qubit eigenbasis goes over to delocalized states, being close to
states that are symmetric and antisymmetric with respect to
reflection around the midpoint of the potential.

This change of the character of the states happens over a very
small interval of $\tilde{\Phi}_c$, because of the exponential
increase of the amplitude of tunneling between the wells as the
barrier between the two wells is decreased. This transition
interval is referred as the ``portal''
\cite{Koch:2005aa,Koch:2006aa}, and pulsing the flux parameters
through this portal can create some of the elementary actions
needed for the construction of quantum gates. For example, a
non-adiabatic pulse through this region can create superpositions
between states $|0\rangle$ and $|1\rangle$.

The fundamental mode of the open-ended transmission line acts as a
harmonic oscillator of frequency $\omega_T$ coupled to the bare
qubit. The presence of that structure modifies the quantum
behavior of the qubit when the energy splitting of the ground and
first excited state of the bare qubit is comparable to or larger
than $\hbar\omega_T$. In this regime, the two lowest-lying states
of the system both have the bare qubit in its ground state; they
differ only in their transmission line quantum number.  By tuning
the energy splitting of the ground and first excited state, one
can move information stored in the bare qubit to the transmission
line and vice-versa.  When the qubit has been transferred to this
transmission-line embodiment, we say that it is {\em parked}.
Parking results in a very useful stabilization of the 0-1
frequency as a function of changes in
$\{\tilde{\Phi}_c,\epsilon\}$. In addition, the quantum coherence
times while parking are seen to reach several $\mu
s$\cite{roger}. Thus, the parking regime will be used as the
memory state of the qubit; it will stay in this state when it is
awaiting operation.  Far away from parking, when the 0-1 energy
gap is much smaller than $\hbar\omega_T$, the transmission line
does not play any role in the dynamics of the 0-1 states, which
are just those of the bare qubit. Since the coherence times here
are expected to be of the order of only tens of $n s$, this regime
should be avoided; this part of parameter space will only be used
for state measurement.

So far, we have given a qualitative description of the qubit dynamics.
This description is made quantitative with the methodology introduced by
Burkard, Koch and DiVincenzo (BKD)\cite{Burkard:2004aa}. Using
network graph theory, BKD developed a universal method for
analyzing any lumped element electrical circuit containing Josephson junctions. The
result of this theory is a mapping of the circuit dynamics to that
of a massive particle in a potential, whose mass tensor and degrees of
freedom are associated with the system capacitances. The system
Hamiltonian in this formulation is given by
\begin{eqnarray}
\fl \cH_S(t)&=&\frac{1}{2}\bQ_C^T\bC^{-1}\bQ_C+\left(\frac{\Phi_0}{2\pi}\right)^2 U(\boldphi,t),\label{BKDhamiltonian}\\
\fl U(\boldphi,t)&=&-\sum_iL_{J;i}^{-1}\cos\varphi_i+\frac{1}{2}\boldphi^T\bM_0\boldphi+\frac{2\pi}{\Phi_0}\boldphi^T[(\bar{\bN}*\tilde{\boldPhi}_x)(t)+(\bar{\bS}*\bI_B)(t)].\label{BKDpotential}
\end{eqnarray}
Here $L_{J;i}^{-1}\equiv\left(\frac{2\pi}{\Phi_0}\right)I_{c;i}$,
where $I_{c;i}$ is the critical current of junction $i$. The
diagonal matrix $\bC$ contains the capacitances of the system. The
topology of the circuit is encoded in the matrices $\bM_0$,
$\bar{\bN}$ and $\bar{\bS}$ (see Appendix for a more detailed
presentation of BKD theory). The first term of the potential Eq. (\ref{BKDpotential})
represents the energy due to the presence of the Josephson junctions.
The second term is associated with the inductive energies of each
branch of the circuit. The last two terms take into account
coupling to external sources of magnetic flux $\tilde{\boldPhi}_x$
and current sources $\bI_B$.  Quantization of the system is
introduced by imposing the canonical commutation relation for the
variables of charge, $\bQ_C$, and phase, $\boldphi$:
\begin{equation}
\left[\frac{\Phi_0}{2\pi}\varphi_i ,
Q_{C;j}\right]=i\hbar\delta_{ij}.
\end{equation}

The analysis of the system potential Eq. (\ref{BKDpotential}) for our qubit reveals that, instead of working with the real applied fluxes $\tilde{\boldPhi}_x\equiv\{\tilde{\Phi}_c,\Phi,\Phi_p\}$, it is more convenient to introduce  ``non-orthogonal'' flux coordinates $\epsilon\equiv\Phi-\Phi_p$ and $\Phi_c\equiv\tilde{\Phi}_c+\frac{L_1-2M_{15}}{2(L_3+M_{35}-M_{15})}(\Phi+\Phi_p)$ (where $L_1$ stands for the small loop inductance, $L_3$ to big loop inductance, and $M_{15}$ and $M_{35}$ to the small-big loop and two big loops mutual inductances, respectively). In fact, as presented in the Appendix, the potential has a definite symmetry as a function of $\{\Phi_c,\epsilon\}$, rather than $\{\tilde{\Phi}_c,\epsilon\}$. Since the potential symmetry is a key feature of system, and we will explore it when designing our gates, an accurate calculation must consider that fact, even though the term proportional to $\frac{L_1-2M_{15}}{2(L_3+M_{35}-M_{15})}\approx0.024$ represents a small correction to the real applied flux $\tilde{\Phi}_c$. From here onwards, we will only refer to the pair of effective fluxes $\{\Phi_c,\epsilon\}$.

Since our qubit has four capacitances (three associated with the
Josephson junctions, and one representing the fundamental mode of
the transmission line), BKD theory leads us to a four-dimensional
potential, for which direct calculations and analysis are more
difficult. Thus, we follow the procedure of
\cite{DiVincenzo:2006aa} to reduce the system dimensionality to
two, one representing the bare qubit and the other representing
the transmission line.  The procedure involves organizing the
degrees of freedom into ``fast'' coordinates, in which the
potential rises very steeply (such that the system dynamics are
frozen into ground state along these direction), and ones that are
``slow''.  Then, using a Born-Oppenheimer approach, the fast
coordinates are traced out, resulting in small modifications to
the remaining slow-coordinate potential energy.

\begin{figure}[t!]
\begin{center}\includegraphics[ width=0.5\columnwidth,
 keepaspectratio]{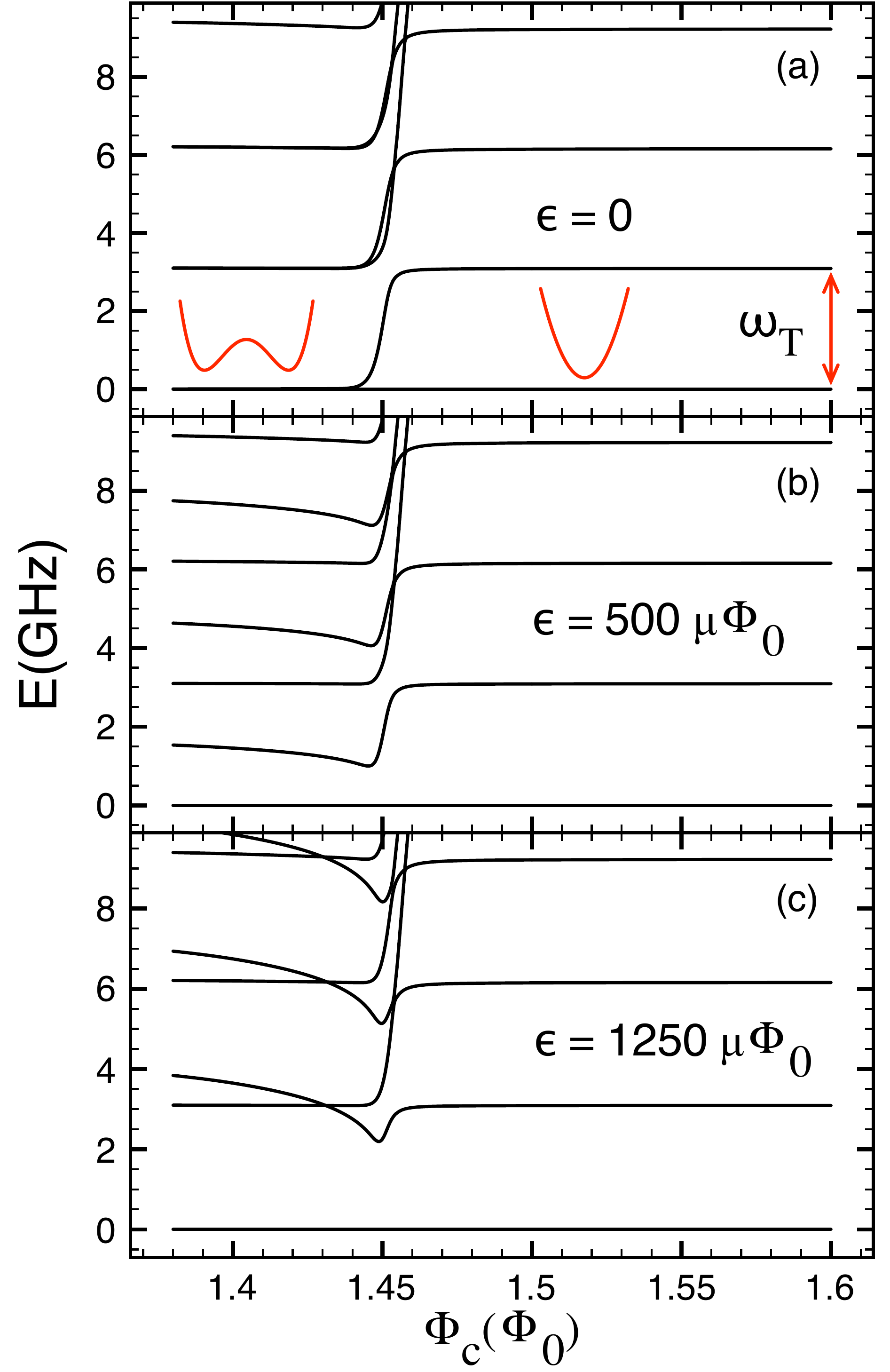}\end{center}
    \caption{The first seven system eigenlevels, as a function of
    the control flux $\Phi_c$, for three different values of the applied
    flux $\epsilon$. The qubit is encoded using the the lowest two eigenstates.
    For large values of $\Phi_c$, the energy splitting of the bare qubit
    states is much larger than $\hbar\omega_T$, so that the lowest
    lying qubit states have purely harmonic oscillator character. (a) The level structure on the S line. In this case,
    the system potential has a symmetric double-well structure for small
    values of control flux, consequently the ground and first excited
    states are nearly degenerate there. (b) and (c) present cases away from the S line. Here, the states
    $|L\rangle$ and $|R\rangle$ are no longer degenerate and the system
    has a gap between the ground and first excited states for small
    values of control flux. }
    \label{spectrum}
\end{figure}

Fig. \ref{spectrum} shows the first seven levels of the system
spectrum, calculated following the above described steps, as a
function of $\Phi_c$, for three different values of the bias flux
$\epsilon$. Note that, by convention, the ground level is always
at $E=0$. It is clear that the equally spaced harmonic oscillator
levels cut through this spectrum for all values of $\Phi_c$ and
$\epsilon$; since the transmission line sees the control flux only
via interaction with the bare qubit, its energies are very stable
except in the vicinity of energy crossings. Observe that the
lowest-lying states in the ``parking" regime, at high values of
$\Phi_c$, involve only transmission line quantum numbers (i.e.,
they are the states of a harmonic oscillator). This is true
because the energy splitting of the eigenstates of the bare qubit
becomes much larger than the $\hbar\omega_T$ energy splitting of
the transmission line states.

In addition, because of the interaction between the bare qubit and
the transmission line, another structure present for each
$\epsilon$ is an observed avoided crossing gap between the bare
qubit and the transmission line states, occurring close to
$\Phi_c=1.45\Phi_0$. As one can see, the bias flux $\epsilon$
plays an important role for small values of $\Phi_c$, where the
system potential has a double-well structure, and the bare qubit
energy splitting is smaller than $\hbar\omega_T$. The first plot
presents the case on the S line, $\epsilon=0$. Since at the S line
the system potential is symmetric, the qubit states are symmetric
and antisymmetric superpositions of the degenerate localized
orbitals of the left and right wells of the potential. As we move
away from the S line ($\epsilon\neq 0$), the symmetry of the
potential is broken and the localized left and right orbitals are
no longer degenerate. This explains the appearance of the gap
between the ground and first excited states in the second and
third plots at small values of $\Phi_c$. One important feature of
the system potential is a symmetry in going from $+\epsilon$ to
$-\epsilon$ (see Appendix); under this transformation, the
energies of the left and right states are interchanged. Finally,
in the third plot it can be seen that the first excited state
stabilizes at the transmission line frequency for $\Phi_c<1.43\Phi_0$.
This occurs because for most values of $\Phi_c$ the bias term in
the Hamiltonian is large enough by itself to make the bare qubit
energy splitting larger than $\hbar\omega_T$ (but not much larger).
 As a result, for small $\Phi_c$ the first excited state
corresponds to one excitation of the transmission line, and the
second excited state to the excited state of the bare qubit.

\section{\label{4level}4-level model}
Although it would be possible, with considerable computational
effort, to design and simulate the desired quantum gates by direct
evaluation of the time dependent Schr\"odinger equation for the
circuit Hamiltonian Eq. (\ref{BKDhamiltonian}), a considerable economy is achieved
by introducing a simplified model that correctly describes the
lowest eigenstates, since we will be only interested in their
dynamics for our quantum gates. We start the derivation of such a simplified model by using first order
perturbation theory to treat the bare qubit coupling with the flux line and with its transmission line. The resulting Hamiltonian due to that approximation has the form of a biased two-level system coupled to a harmonic oscillator (see Appendix for a detailed derivation)
\begin{eqnarray}
H=-\frac{1}{2}\Delta(\Phi_c)\hat{\sigma}_x+\frac{1}{2}\epsilon b(\Phi_c) \hat{\sigma}_z {}
+\hbar\omega_{T} \hat{a}^\dagger \hat{a}+g(\Phi_c)(\hat{a}+\hat{a}^\dagger)\hat{\sigma}_z,
\label{hamiltonianapp}
\end{eqnarray}
where $\hat{\sigma}_i$ are the Pauli matrices, $\hat{\sigma}_z|L\rangle\equiv-|L\rangle,~\hat{\sigma}_z|R\rangle\equiv+|R\rangle$, and $\{\hat{a}^\dagger,\hat{a}\}$ are canonical bosonic creation and annihilation operators. The states $|L\rangle$ and $|R\rangle$ represent the localized orbitals found in the left and right potential wells.

\begin{figure}[t!]
\begin{center}\includegraphics[ width=0.5\columnwidth,
 keepaspectratio]{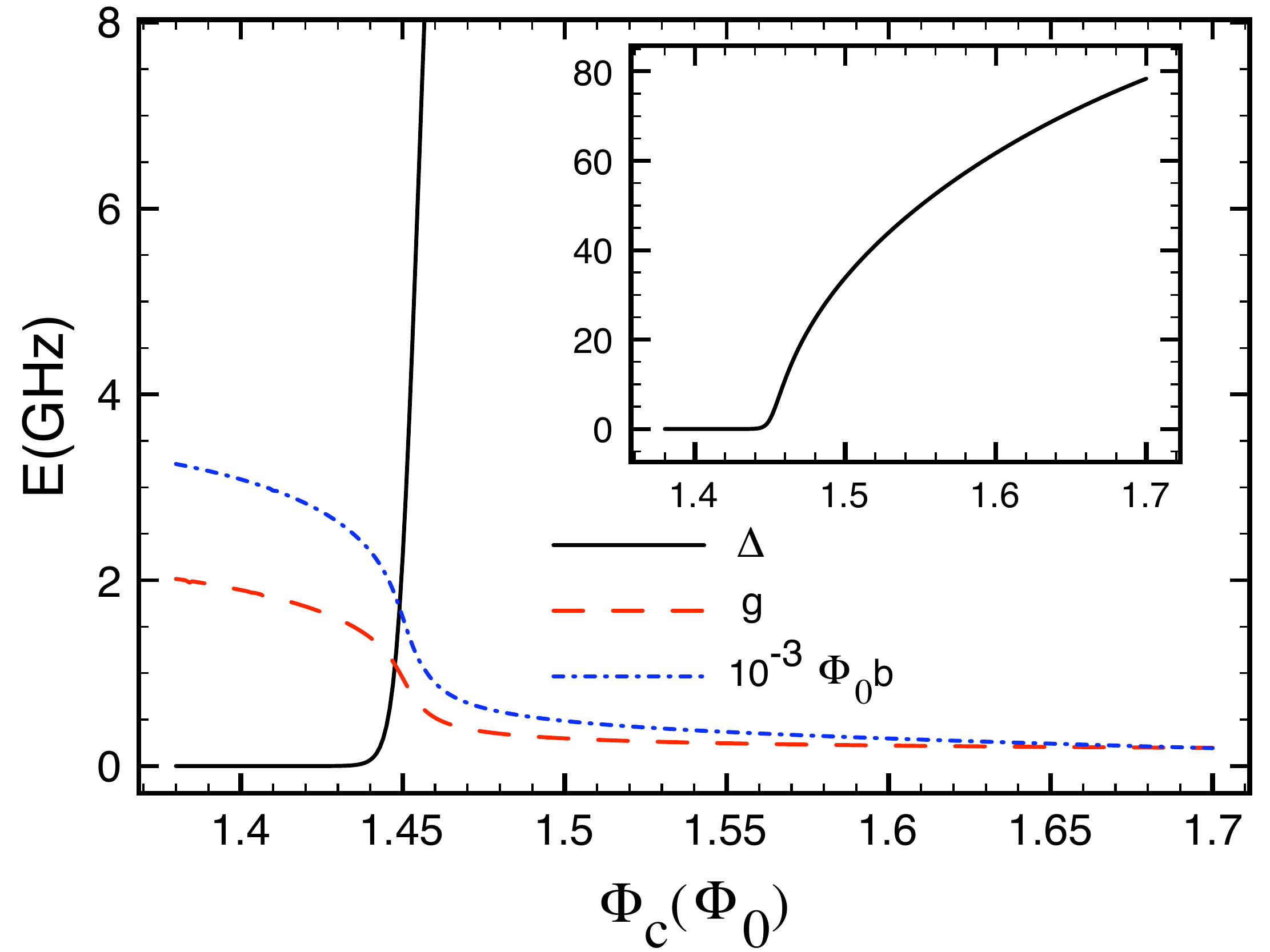}\end{center}
    \caption{Coefficients of the one-qubit Hamiltonian Eq. (\ref{hamiltonianapp})
    as a function of the control flux $\Phi_c$. The solid curve
    gives the tunneling amplitude between the ground and
    first excited states of the system. The dashed
    curve gives the coefficient of the coupling between the bare qubit and the transmission
    line, and dot-dashed curve gives the bias term. For small values of
    $\Phi_c$, the tunneling amplitude becomes exponentially
    suppressed due to the very large barrier between the two wells of the
    potential. As one increases $\Phi_c$, the potential
    barrier decreases and the positions of the minima become closer together.
    Around $\Phi_c=1.45\Phi_0$ the barrier vanishes entirely; from
    this point onward, the potential has a
    single well, as sketched in Fig. \ref{spectrum}. Inset: detail of the tunneling amplitude for the single-well regime.}
    \label{hamiltonian}
\end{figure}

Because of the change of the state character as a function of $\Phi_c$, the amplitude of tunneling $\Delta$, the bias term coefficient $b$ and the qubit-transmission line coupling $g$ also become control flux dependent. Fig. \ref{hamiltonian} presents their behavior as a function of $\Phi_c$. As one can see, the tunneling amplitude $\Delta$ becomes negligible for small values of $\Phi_c$. This happens because, in this regime, the very large barrier between the two wells ($\gg100$GHz) in the double-well potential exponentially suppresses the tunneling between their lowest states. However, as the value of $\Phi_c$ increases, the barrier decreases and the two minima become closer. Consequently, the amplitude of tunneling rapidly increases, and the left and right orbitals start to become more and more delocalized. Around $\Phi_c\approx1.45\Phi_0$ the barrier vanishes rather abruptly and the bare qubit potential enters a single-well regime. The region around $\Phi_c\approx1.45\Phi_0$ defines the portal of the system. The dashed and dot-dashed curves present the coefficient of the bare qubit-transmission line coupling $g$, Eq. (\ref{gequation}), and the bias term $b$, Eq. (\ref{bequation}), respectively. Once the bare qubit reaches the single well regime, those coefficients become less sensitive to changes in $\Phi_c$, since the nature of the states barely changes as one passes through this regime. In addition, the inset shows the amplitude of tunneling can rapidly reach values of several tens of GHz in the parking regime, which are, in general, much larger than the typical value of the transmission line energy splitting $\hbar\omega_T$, hence the lowest-lying states in the parking regime involve only transmission line states.

\begin{figure}[t!]
\begin{center}\includegraphics[ width=0.5\columnwidth,
 keepaspectratio]{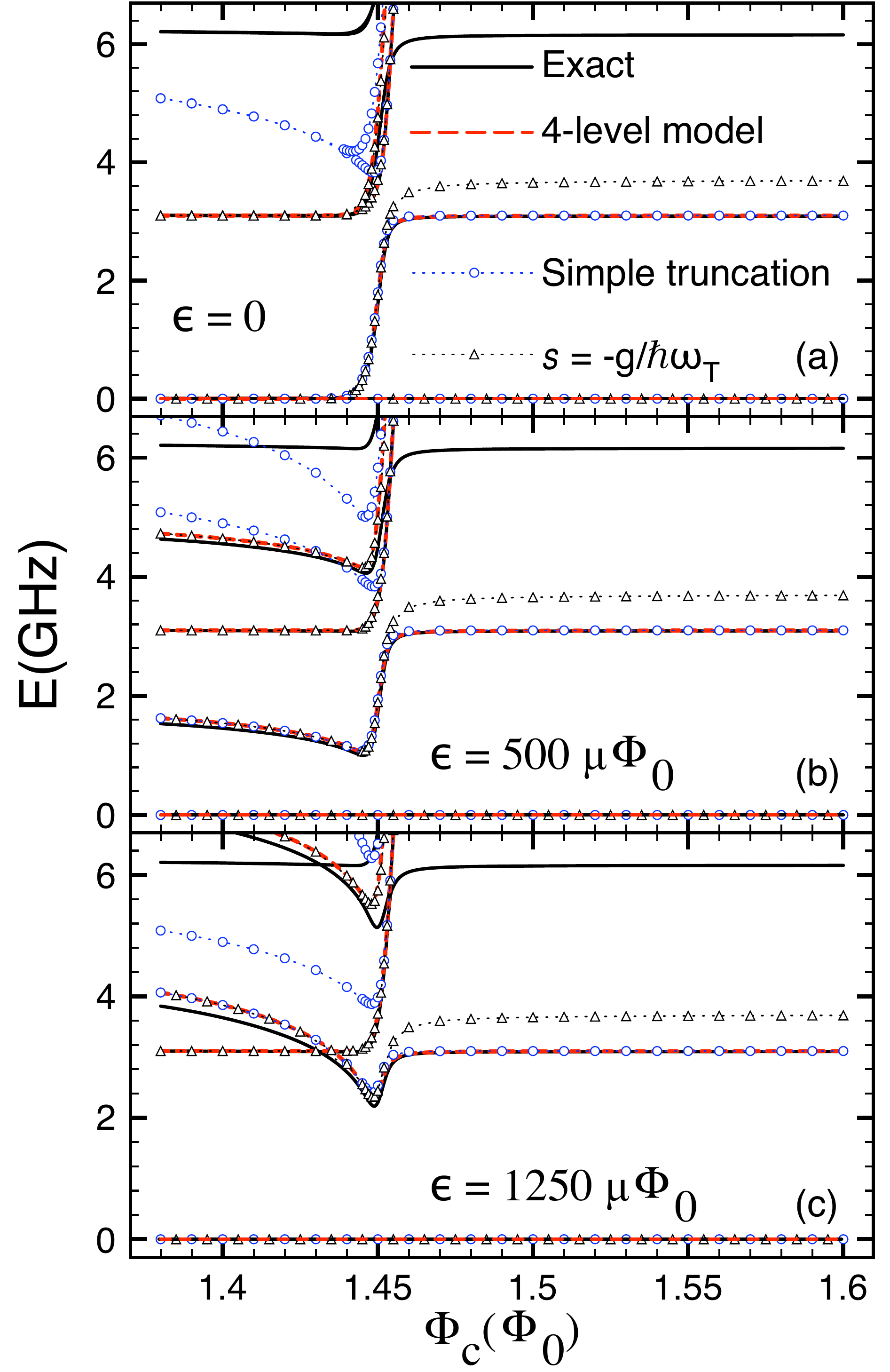}\end{center}
    \caption{The first system eigenstates as a function of control flux $\Phi_c$, obtained through the Schr\"odinger equation solution of the exact potential Eq. (\ref{BKDpotential}) (solid curve), a simple truncation of Hamiltonian Eq.(\ref{hamiltonianapp}) at the first excited state of harmonic oscillator (circle-symbolized curve), and the solution of the 4-level proposed, Eq. (\ref{4model}), using the shifts $s=-g/\hbar\omega_T$ (triangle-symbolized curve) and $s=-g/\sqrt{\omega_T^2+\Delta^2}$ (dashed red curve), Eq. (\ref{shift}), for three different values of the bias flux $\epsilon$. The simple truncation fails when $g$ becomes appreciable (small values of $\Phi_c$). In this regime, the shifted harmonic oscillator states are the preferred system representation. The 4-level model using the shift $s=-g/\hbar\omega_T$ does not provide the parking harmonic stabilization, since it does not consider the change of character of the bare qubit. At last, the 4-level model using the conditional shift Eq. (\ref{shift}) gives a very fair approximation of the lowest-lying levels for all regimes of the bias and control flux.}
    \label{4by4}
\end{figure}

Unfortunately, even though the Hamiltonian Eq. (\ref{hamiltonianapp}) already represents an important simplification for the system description, since it has reduced the bare qubit Hilbert space to that of a two-state system, we cannot obtain an analytical solution for its eigenstates/eigenvalues. Thus, we still have to deal with an infinite set of states due to the harmonic oscillator. As we shall see, for the purposes of our work, we must have not only the correct dynamics of the states $|0\rangle$ and $|1\rangle$, but also an excellent agreement for the minimal gap between the computational basis $\{|0\rangle,|1\rangle\}$ and the rest of spectrum of the system. Because of that, the truncation of the harmonic oscillator Hilbert space at its first excited state does not work as a fair approximation of our system Hamiltonian. As presented in Fig. \ref{4by4}, the simple truncation of the harmonic oscillator Hilbert space fails to give a good description of the gap between the $|1\rangle$ and $|2\rangle$ states. This happens because at the regime of small values of $\Phi_c$ the function $g$ is appreciable. Hence, the shift applied to the oscillator due to the interaction with the bare qubit becomes important. Consequently, the most adequate description of the system is obtained using the representation of shifted harmonic oscillator states, where the truncation of the harmonic oscillator Hilbert space should give better results.

In fact, a more careful inspection of the Hamiltonian Eq. (\ref{hamiltonianapp}) reveals that its harmonic oscillator part has a form of a well-known shifted oscillator. Thereby, changing to the representation of the shifted harmonic oscillator states should give us a better picture of the system dynamics, and the appropriate basis in order to perform further truncations of the Hilbert space. Nevertheless, the shift applied to the harmonic oscillator is dependent on the  bare qubit state, as can be seen by the term $g(\Phi_c)(\hat{a}+\hat{a}^\dagger)\hat{\sigma}_z$ of Eq. (\ref{hamiltonianapp}). This important feature of the system leads us to introduce the unitary transformation $\hat{D}(s,\hat{\sigma}_z)\equiv e^{(s\hat{a}^\dagger-s^\ast\hat{a})\hat{\sigma}_z}$, as the conditional displacement operator of the system. Observe that the operator $\hat{D}$ does not commute with the spin part of the Hamiltonian Eq. (\ref{hamiltonianapp}). Thus, changing the Hamiltonian representation to the shifted harmonic oscillator states, $H\rightarrow \hat{D}^\dagger H \hat{D}$, and then performing the truncation of the harmonic oscillator Hilbert space at its first excited state, we arrive with the following simplified 4-level model

\begin{eqnarray}
H_{4-level}=&&-\frac{1}{2}\Delta(\Phi_c)e^{-2|s|^2}e^{-2s\hat{\sigma}_z\hat{c}^\dagger}\hat{\sigma}_x e^{-2s^\ast\hat{c}\hat{\sigma}_z} +\frac{1}{2}\epsilon b(\Phi_c)\hat{\sigma}_z \nonumber\\
& &+\hbar\omega_T\hat{c}^\dagger\hat{c}+\hat{\sigma}_z\hat{c}(g(\Phi_c)+s^\ast\hbar\omega_T)+\hat{c}^\dagger\hat{\sigma}_z(g(\Phi_c)+s\hbar\omega_T),
\label{4model}
\end{eqnarray}
where we have defined the operators $\hat{c}\equiv|0_{{\rm HO}}\rangle\langle1_{{\rm HO}}|$ and $\hat{c}^\dagger\equiv|1_{{\rm HO}}\rangle\langle0_{{\rm HO}}|$, in which $|0_{{\rm HO}}\rangle$ and $|1_{{\rm HO}}\rangle$ represent the ground and first excited harmonic oscillator states, respectively. During the procedure described above, we have introduced an {\it ad hoc} parameter, the shift parameter $s$, which can be used to parametrize the Hamiltonian Eq. (\ref{4model}) in order to obtain the closest level dynamics as possible to that determined by the Hamiltonian Eq. (\ref{BKDhamiltonian}).

A natural choice for the parameter $s$ would be the standard value $s=-g(\Phi_c)/\hbar\omega_T$, which leads to the cancellation of the last two terms of Eq. (\ref{4model}), and the Hamiltonian diagonalization when $\Delta\approx0$. However, since the shift imposed on the harmonic oscillator is conditioned on that of the bare qubit state, and the form chosen for $s$ does not take into account the change of character of the bare qubit state, one should expect it would fail when the regime of high tunneling amplitude is reached. Indeed, as can be seen in Fig. \ref{4by4}, this choice for the parameter $s$ only gives the correct description when the tunneling amplitude is negligible. In fact, the analytical solution of Eq. (\ref{4model}) using the above shift reveals that, at the limit $\Delta\rightarrow\infty$, one should expect $\omega_{01}\rightarrow\infty$, what is in complete opposition with the parking stabilization expected for the system eigenstates.

Therefore, a smart choice of the parameter $s$ has to consider the fact of the harmonic oscillator shift is conditioned to the bare qubit state, and that its state character changes as a function of $\Phi_c$. As we already know, at the limit $\Delta\rightarrow0$, the parameter $s$ should asymptotically reach the value $s\rightarrow-g(\Phi_c)/\hbar\omega_T$, since it correctly decouples the spin and harmonic oscillator degrees of freedom in the Hamiltonian Eq. (\ref{4model}). In the other limit, $\Delta\rightarrow\infty$, one should expect observing no shifts to be imposed on the harmonic oscillator, since the system states would be frozen at the bare qubit ground state. Thus, we expect a reasonable interpolation for the parameter $s$ between the two regimes is
\begin{equation}
s(\Phi_c)=-\frac{g(\Phi_c)}{\sqrt{\omega_T^2+\Delta^2(\Phi_c)}}.
\label{shift}
\end{equation}

In fact, as presented by the dashed curve in Fig. \ref{4by4}, the form Eq. (\ref{shift}) for the parameter $s$ gives a very fair approximation for the lowest level dynamics, in particular the energy splittings $\omega_{01}$ and $\omega_{12}$. In addition, from the analytical solution of Eq. (\ref{4model}) using Eq. (\ref{shift}), we can check that the $\omega_{01}$ has the correct harmonic oscillator stabilization, {\it i. e.} $\Delta\rightarrow\infty\Rightarrow\omega_{01}\rightarrow\omega_T$.

It is worth pointing out that, although the model Eq. (\ref{4model}) does not give the correct parking stabilization for the system second excited state $|2\rangle$, since this state involves the second excited state of the transmission line, for the purposes of our work, it turns out that is not a limitation for the model. Indeed, the lack of parking stabilization for the second excited state $|2\rangle$ would lead to underestimations of the transitions between the state $|1\rangle$ and $|2\rangle$, and consequently wrong leakage estimations. However, as we are going to assume $\omega_T$ of the order of several GHzs, the energy splitting $\omega_{12}$ for the parking regime is big enough to suppress those transitions during our dc shaped pulse operations. Furthermore, since at the parking regime the ground state, and the first and second excited states have the same nature ({\it i. e.} they are the states of the harmonic oscillator), observing the induced 0-1 transitions due to the dc pulse operations also gives a measurement of the expected 1-2 transitions in this regime. Thus, we end with a very controllable four dimensional Hilbert space model, spanned by the ground and first excited states of the bare qubit and transmission line, that accurately mimics the main features of the system dynamics determined by the exact circuit Hamiltonian model Eq. (\ref{BKDhamiltonian}). From now onwards, we will only use the Hamiltonian Eq. (\ref{4model}), with Eq. (\ref{shift}), when designing and simulating the quantum gates.

\begin{figure}[t!]
\begin{center}\includegraphics[ width=0.5\columnwidth,
 keepaspectratio]{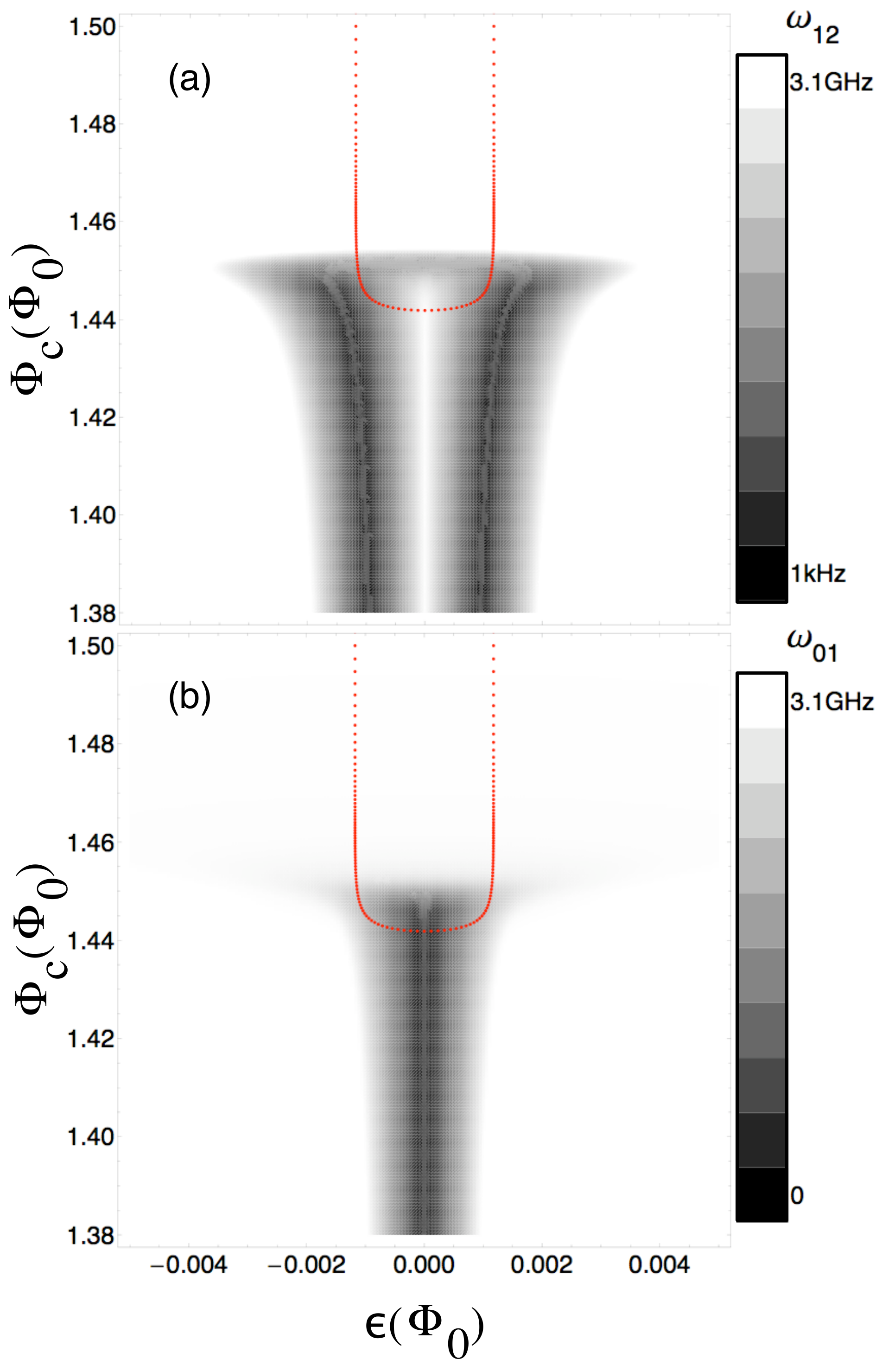}\end{center}
    \caption{Plots of the energy splittings (a) $\omega_{12}$ and (b) $\omega_{01}$
    for the flux space $\{\Phi_c,\epsilon\}$. The plots provide a
    good ``map'' for designing the one-qubit gates, since they
    indicate the regimes where one should expect an appreciable
    amount of leakage and very fast phase accumulation. (a) The
    $\omega_{12}$ plot quantifies the minimal gap between the
    computational basis $\{|0\rangle,|1\rangle\}$ and the rest
    of spectrum of the system, giving a limit on the rate at which the system can evolve without
    appreciable leakage. (b) The $\omega_{01}$
    plot indicates the phase accumulation rate with respect to the reference. In regions of very fast
    rates, a very precise control of the external applied fluxes is
    required in order to avoid phase noise. Illustrated with dots is the path in the flux space used to
    implement the Hadamard gate. The number of dots indicates the time spent when passing through
    that region.}
    \label{densitplots}
\end{figure}

Completing the system description, Fig. \ref{densitplots} presents the plots for the frequency difference between the
ground and first excited state, $\omega_{01}$, and the frequency
difference between the first two excited states, $\omega_{12}$, as
a function of $\{\Phi_c,\epsilon\}$.  As mentioned before, the $\omega_{12}$ plot
quantifies the minimal gap between the computational basis
$\{|0\rangle,|1\rangle\}$ and the rest of spectrum of the system,
giving a measure of how likely the system is to undergo leakage
during the sweep of a pulsed gate.  The $\omega_{01}$ plot shows
how fast the relative phase of the two computational basis states
advances if the system is held at some flux values.  We envision
that there will be a very stable master clock at the
transmission-line frequency, and that errors in the accumulated
phase difference are unlikely if the system is held in parking,
where its phase advance is synchronous with this clock.  But when
$\omega_{01}$ departs far from $\omega_T$, phase accumulation with
respect to the reference is very fast, and we expect (and our
numerical studies confirm) that the system is much more
susceptible to phase errors in this regime. These plots provide a
good ``map'' for designing the one-qubit gates, since they
indicate the regimes where one should expect an appreciable amount of
leakage and very fast relative phase accumulation.

\section{\label{gates}One-qubit Gates}

Before we move on to the discussion of each gate individually, it is worth summarizing some features present in all of them (including the two-qubit gate). First the memory and measurement points must be defined. The parking point is taken to be at $\Phi_c=1.6\Phi_0$, and the
measurement point to be at $\Phi_c=1.4\Phi_0$. These choices are
optimal, in the following sense: For the parking point, the
important feature to be considered is how much the $\omega_{01}$
frequency changes as a function of external controls. Since we
envision the qubit spending long times at this position, it is
imperative that the frequency deviation $\delta\omega_{01}$ have
very small values for reasonable flux shifts due to the noise. We
calculate that the frequency sensitivity is
$\delta\omega_{01}/\delta\Phi_c\approx 10\rm{MHz}/\Phi_0$ at the chosen memory point, which
will give acceptably small memory error. In addition, inaccuracies in the transmission line fundamental model frequency can be corrected using a $\pi$-pulse stabilization scheme \cite{PhysRev.80.580}. This operation (not analyzed in detail here) consists in initially applying a $\pi$ rotation gate; the system is then left to evolve in a free evolution for the same amount of time previously spent at the memory state, and finally a new $\pi$ rotation is applied. As a result, undesired Z-phase accumulated at the parking point ends as a global phase of the system state.

For the measurement
point, we must make sure the left and right states are
experimentally distinguishable, and that, since the potential has a
periodic structure \cite{DiVincenzo:2006aa}, the potential barrier between
different pairs of minima is high enough to avoid tunneling
between their states. At $\Phi_c=1.4\Phi_0$ we found the barrier
between the two principal minima is of the order of 10THz, and
the barrier to other minima is even higher (c. 20THz).

Another common feature for all of our shaped pulses is our design methodology. We shape our dc pulses using simple sums of $\tanh$ functions. As we shall see, our gates can be divided in several parts, and with each part is associated a function of the form $\delta\tanh\left(\frac{t-t_{max}}{\tau}\right)$, where $\delta$, $\tau$ and $t_{max}$ determine the flux excursion, the maximum rate and its time position, respectively. Thus, with those parameters, one can adjust the rate and the flux excursion of each part of the gate, in order to optimize the gate. It turns out that our designed gates do not require maximum flux slew rate and bandwidth higher than $7\times10^6\Phi_0/$s and $1$GHz, respectively.

We measure the gates' fidelity using the entanglement fidelity \cite{Schumacher:1996aa}
\begin{equation}
F=(\rm{Tr}\rho^Q A^Q)(\rm{Tr}\rho^Q A^{Q\dagger}),\label{fidelity}
\end{equation}
where ${\rm A^Q}\equiv U_{ideal}^\dagger U_{real}(\delta\Phi_c,\delta\epsilon,\delta t)$, with $U_{ideal}$ and $U_{real}(\delta\Phi_c,\delta\epsilon,\delta t)$ representing the ideal gate and the final achieved transformation, respectively. $\rho^{Q}$ is an equal distribution of the computational basis states $|0\rangle$ and $|1\rangle$: $\rho^{\rm Q}\equiv\frac{1}{2}|0\rangle\langle0|+\frac{1}{2}|1\rangle\langle1|$. Finally, we estimate the probability of leakage using the projection of an arbitrary evolved state, $|\Psi_f\rangle=U|\Psi_i\rangle$, in the 0-1 computational basis: $1-\langle\Psi_f|(P_0+P_1)|\Psi_f\rangle=1-|(P_0+P_1)U|\Psi_i\rangle|^2$, where $P_j$ represents the projection operator in the state $|j\rangle$.

Figures \ref{measurement}, \ref{zgate} and \ref{hadamard} show the
proposed measurement gates, phase gate, and the Hadamard
gate, whose matrix representation in the $\{|0\rangle,|1\rangle\}$ basis are respectively given by
\begin{equation}
U_{0/1}=\left(\begin{array}{cc}
1&0\\
0& e^{i\theta_{01}}
\end{array}\right),\label{01}
\end{equation}

\begin{equation}
U_{+/-}=\frac{1}{\sqrt{2}}\left(\begin{array}{cc}
1&1\\
e^{i\theta_{+-}}& -e^{i\theta_{+-}}
\end{array}\right),\label{+-}
\end{equation}

\begin{equation}
U_{phase}=\left(\begin{array}{cc}
1&0\\
0& e^{i\theta_{z}}
\end{array}\right),\label{phase}
\end{equation}

\begin{equation}
U_{Hadamard}=\frac{1}{\sqrt{2}}\left(\begin{array}{cc}
1&1\\
1& -1
\end{array}\right).\label{hadamardmatrix}
\end{equation}
Where $\theta_{01}$ and $\theta_{+-}$ are arbitrary phases, whose
values are not relevant for the gate implementation, provided that
the gates Eqs. (\ref{01}) and (\ref{+-}) are followed by a
projection into the basis $\{|0\rangle,|1\rangle\}$. The relative
phase $\theta_z$ is the parameter that defines the phase gate, Eq.
(\ref{phase}). Finally, since the states $|0\rangle$ and
$e^{i\theta_0}|0\rangle$, and  $|1\rangle$ and
$e^{i\theta_1}|1\rangle$ are physically identical, the convention
for the global phases $\theta_0$ and $\theta_1$ is chosen such
that the physical implementation of the Hadamard gate has the
matrix representation given by Eq. (\ref{hadamardmatrix}). Once
fixed, the phase convention is maintained for all other gate
implementations.

In addition, figures \ref{measurement}, \ref{zgate} and
\ref{hadamard} also present the gate fidelities as a function of
unwanted constant shifts from the optimal point of operation in
the whole pulse profile.  We model the effect of 1/f flux noise by
an ensemble of random constant shifts of this sort.  Our
assumption of constancy is accurate for those components of the
1/f noise at frequencies below the inverse of the gate time,
around 100MHz.  1/f noise at this frequency and higher is not
accurately modelled in this way, but at these frequencies we
believe that other sources of high frequency noise (that is, white
noise from the resistances in our circuit) become more important
than 1/f noise. In our work these noise sources are modelled
separately using a quantum bath; see Ref.~\cite{DiVincenzo:2006aa}
for these calculations.

Our estimates for $T_1$ and $T_2$ \cite{Burkard:2004aa} times
indicate that we can expect a very long coherence time in parking,
$O(1s)$, and a very short dephasing time, $~10\rm{ns}$, at the
measurement position. In the portal region, we calculate coherence
times of the order of hundreds of $\mu s$. As we shall see, except
in the measurement processes, our gates are designed to not go
lower than the upper limit of the portal. Thus, in principle, for
gates of duration times of 20-30ns, the fidelity should not be
compromised by decoherence much more strongly than by the
imperfections explored in these figures: unwanted shifts in the
applied fluxes due to low-frequency noise, and temporal shifts
between $\Phi_c$ and $\epsilon$ pulses.

Those shifts are assumed uncorrelated and constant during a single
gate operation, but random from one ``shot" to another. We model
the noise as a normalized gaussian probability distribution,
$\mu(x)\equiv\frac{1}{N}e^{-x^2/2\sigma^2}$, which we assume
having (at 1Hz) $6\mu\Phi_0$ (flux shifts) and $6\rm{ps}$ (time
shifts) as its root-mean-square deviation, $\sigma$, such that
approximately $90\%$ of the distribution is found between the
values $\pm10\mu\Phi_0$ (flux shifts) and $\pm10\rm{ps}$ (time
shifts); we believe that this quality of control will be readily
achievable in the lab in the coming years. Indeed, recent
experimental results \cite{bialczak} indicate that our assumption
of having $6\mu\Phi_0$ as root-mean-square deviation of the 1/f
flux noise at 1Hz is already achievable for Josephson qubits.

We note that by using the 1/f rms amplitude at 1Hz in this model,
we are implicitly assuming that the 1/f noise is cut off below
this frequency.  In fact, the 1/f spectrum goes much lower (at
least three orders of magnitude lower in Ref.~\cite{bialczak}, and
in other similar previous studies); but we may assume that in
quantum computer operation, qubits are frequently taken off-line
and recalibrated. Assuming that this recalibration takes place
once per second gives the 1Hz cutoff. Our assumption of Gaussian
statistics merely embodies the expectation that the 1/f noise
arises as the summed effect of many independent fluctuators; this
is also borne out by the traces of Ref.~\cite{bialczak}.

Table \ref{table} summarizes the average fidelity obtained for the
gates discussed considering each channel noise separately:
$\langle F\rangle=\int d\delta x\mu(\delta x) F(\delta x)$, where
$F$ is given by Eq. (\ref{fidelity}). As one can see, the minimal
expected fidelity found was $99.47\%$ for the Hadamard gate.

For the phase, Hadamard and controlled-Z gates, we present the operator-sum representation\cite{nielsenchuang} of the system superoperator
\begin{eqnarray}
{\cal E}(\rho)&=&\int d\delta \vec{x}\mu(\delta \vec{x}) U_{real}(\delta \vec{x})\rho U_{real}^\dagger(\delta \vec{x})\nonumber\\
&\equiv&\sum_k E_k\rho E_k^\dagger,\label{operatorsum}
\end{eqnarray}
where the operators $\{E_k\}$ are the operation elements for the quantum operation $\cal E$. Having the set of operators $\{E_k\}$ permits the characterization of the noise present during the system evolution. We find that the model noise considered in this work indicates that the effective noise for the physical implementation of our universal set of quantum gates is heavily biased, {\it i. e.}, we have found that the effect of  phase noise is at least one order of magnitude higher than bit-flip noise and leakage processes.  New strategies for fault tolerant computation have recently been worked out which take advantage of this sort of biassing in the noise \cite{Panos}.

\begin{table}
\caption{\label{table}The expected gate fidelities, as a function of unwanted shifts in $\delta\Phi_c$, $\delta\epsilon$ and $\delta t$, calculated considering each channel noise separately. In addition, the expected total net fidelity evaluated in the presence all channels is presented.}
\begin{indented}
\item[]\begin{tabular}{|c|c|c|c|c|c|}
\hline
\multicolumn{6}{|c|}{One-qubit system}\\
\hline
Gate& & $\delta\Phi_c$ & $\delta\epsilon$ & $\delta t$ & All\\
\hline
0/1 Measurement& $\langle\rm{Fidelity}\rangle(\%)$& $99.99$ & $99.99$ & - & $99.99$\\\hline
+/- Measurement & $\langle\rm{Fidelity}\rangle(\%)$& $99.81$ & $99.84$ &- & $99.8$\\\hline
Phase Gate & $\langle\rm{Fidelity}\rangle(\%)$& $99.999$ & $99.999$ &- &$99.999$\\ \hline
Hadamard & $\langle\rm{Fidelity}\rangle(\%)$& $99.87$ & $99.47$ & $99.63$ & $99.46$\\\hline
\end{tabular}
\end{indented}
\end{table}

\subsection{Measurement gates}

\begin{figure}[t!]
\begin{center}\includegraphics[ width=0.5\columnwidth,
 keepaspectratio]{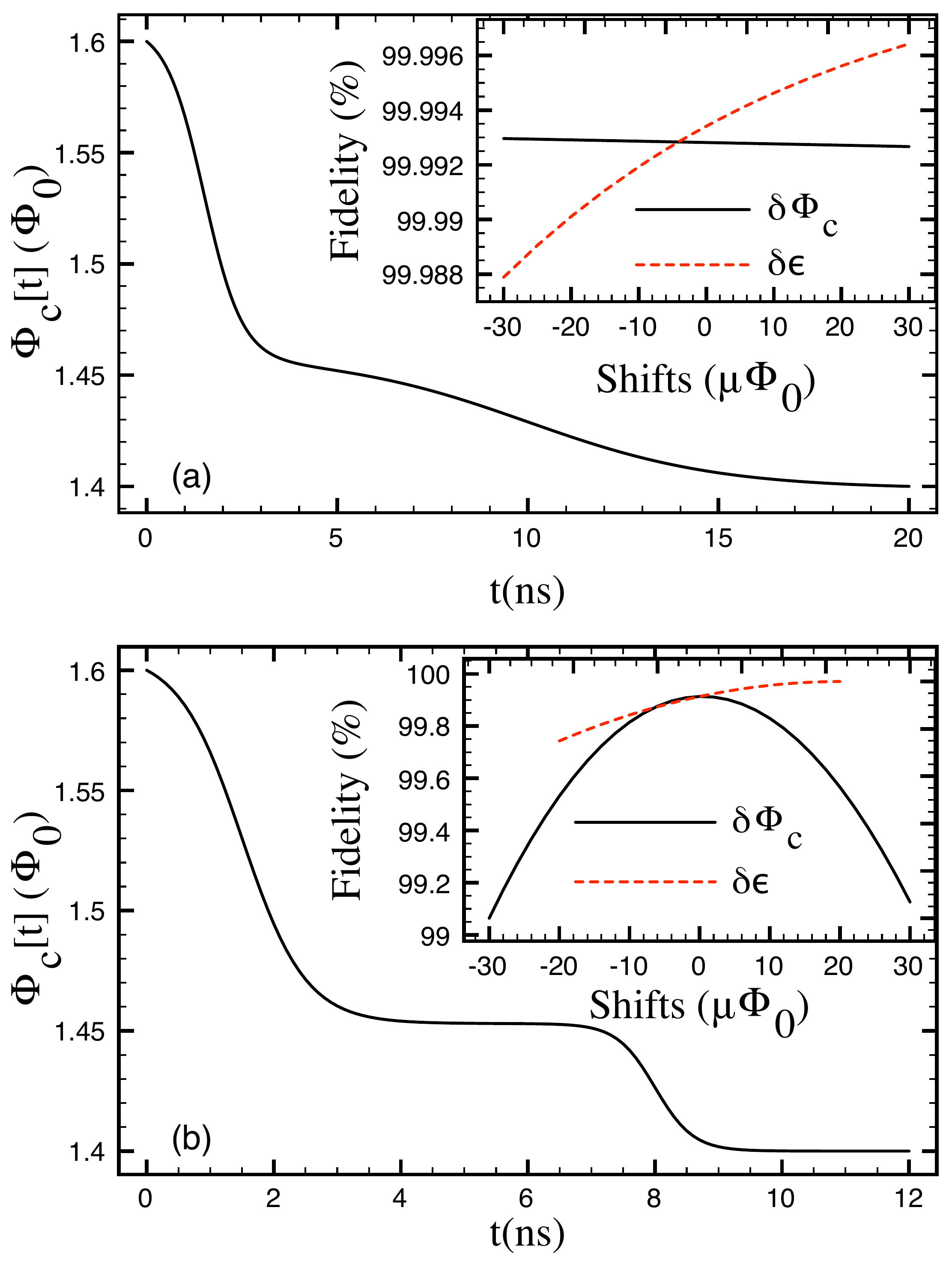}\end{center}
    \caption{Measurement gates. (a) The $\Phi_c$
    pulse path used to perform $0/1$ measurement. The gate is designed
    to bring the system from the parking regime to the measurement
    point through an adiabatic process.  The applied flux $\epsilon$
    is maintained constant during the gate, and distant from the S line:
    $\epsilon=500\mu\Phi_0$. (b) The $+/-$
    measurement gate. The pulse evolves the system from the
    parking regime to the measurement point, performing the unitary
    transformation
    $|+\rangle=\frac{1}{\sqrt{2}}(|0\rangle+|1\rangle)\rightarrow |R\rangle$
    and $|-\rangle=\frac{1}{\sqrt{2}}(|0\rangle-|1\rangle)\rightarrow |L\rangle$.
    The gate consists of two different regimes: the first,
    $t\lesssim 7\rm{ns}$, is designed to be an adiabatic process,
    thus minimizing leakage when passing through the avoided crossing.
    The other, $t>7\rm{ns}$, evolves the system in a non-adiabatic
    process through the portal, leading to the desired transitions
    between $|0\rangle$ and $|1\rangle$ states. The applied flux
    $\epsilon$ is maintained constant during the gate, and close
    to the S line: $\epsilon=30\mu\Phi_0$. Insets: The gate fidelity
    as a function of unwanted shifts in $\Phi_c$ and $\epsilon$
    from the optimal point of operation. The main source for loss
    of fidelity for the $0/1$ measurement is related with $0/1$
    transitions, and for the $+/-$ measurement it is related to
    the phase noise. Note the difference in the scales.}
    \label{measurement}
\end{figure}

\begin{figure}[t!]
\begin{center}\includegraphics[ width=0.5\columnwidth,
 keepaspectratio]{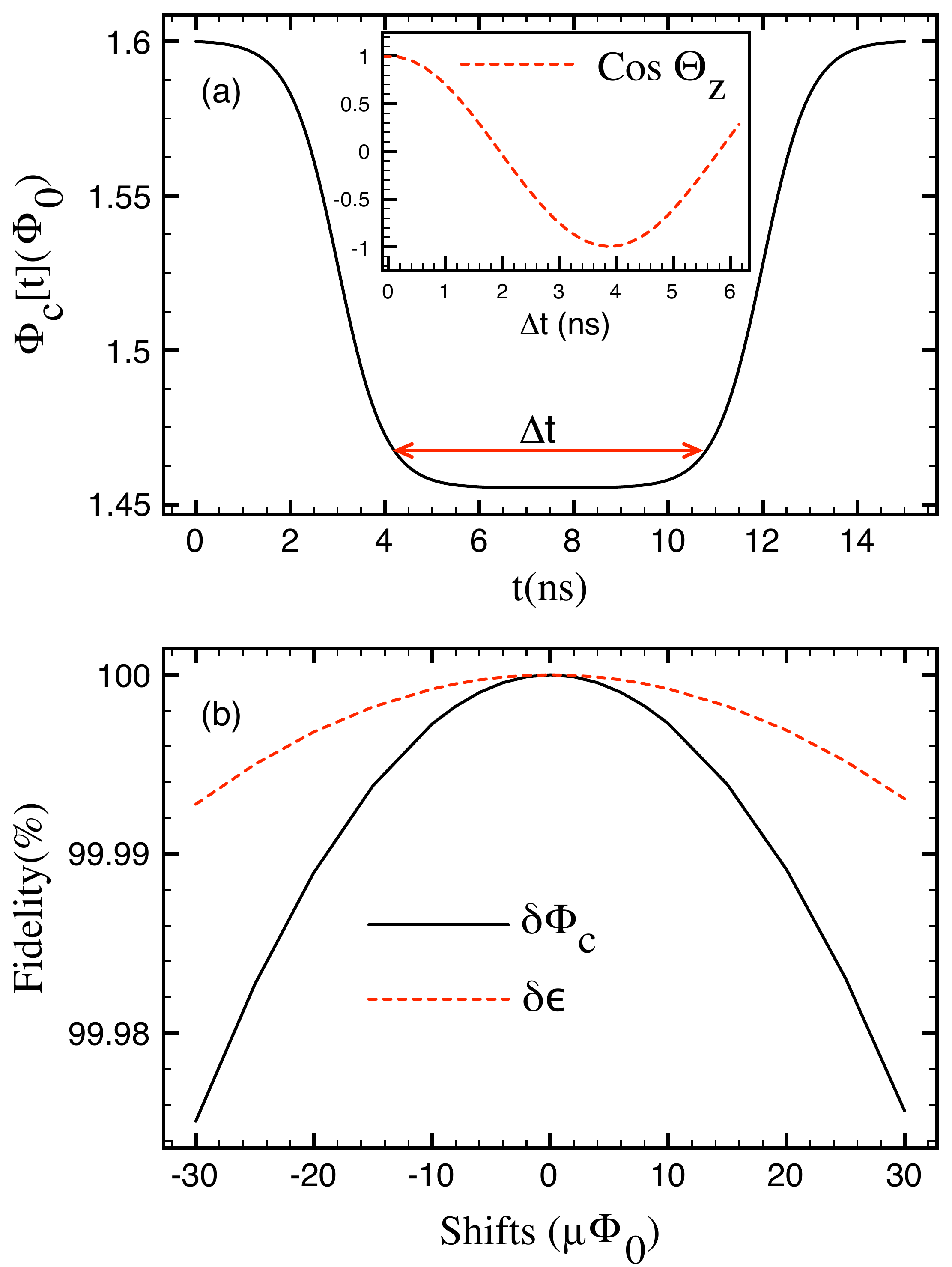}\end{center}
    \caption{The phase gate. (a) The gate is designed
    to perform an adiabatic process, evolving the system from
    the parking regime to a position where the phase rate
    $\omega_{01}$ becomes appreciable compared with the reference.
    As one can see, that position can be reached above the
    avoided crossing gap, which greatly minimizes leakage
    for this operation. After the appropriate amount of time
    $\Delta t$ at the bottom of the pulse (inset presents
    $\cos\theta_z$ as a function of $\Delta t$), the system is
    brought back to the initial position. The flux $\epsilon$
    is kept constant during the operation. (b) The fidelity of the gate as a function of unwanted shifts in
    $\Phi_c$ and $\epsilon$ from the optimal point of operation.
    The main source for the loss of fidelity is related with phase
    noise.}
    \label{zgate}
\end{figure}

We present two distinct measurement gates, both very useful for
effective quantum error correction and universal quantum
computation. As one can see at their matrix representations Eqs.
(\ref{01}) and (\ref{+-}), by measurement gates we mean the
unitary operations that are used to prepare the state for the
final projection process (measurement). The first gate is a
measurement in the standard 0/1 basis. That is, starting from
parking, we are to distinguish whether the system is in its ground
state (0 quanta in the transmission line mode) or its first
excited state (1 quantum in the transmission line mode). This
gate, shown as the first plot of Fig. \ref{measurement}, works by
performing an adiabatic evolution of the states $|0\rangle$ and
$|1\rangle$ from the parking point to the measurement point,
maintaining the bias flux $\epsilon$ at a constant value off the S
line $(\epsilon\neq0)$. So, through this adiabatic transformation,
the qubit states evolve from the configuration $|S,0\rangle$
(ground state) and $|S,1\rangle$ (first excited state), to
$|L,0\rangle$ and $|R,0\rangle$ (where the first label corresponds
to the bare qubit states, and the second to those of the
transmission line). The matrix representation Eq. (\ref{01}) shows
that we end with an irrelevant relative phase $\theta_{01}$
between the $|0\rangle$ and $|1\rangle$ states. The sources of
loss of fidelity are leakage at the avoided crossing gap, and 0/1
transitions at the portal. For the path shown, the probabilities
of observing leakage and 0/1 transitions are $3\times10^{-4}\%$
and $6\times10^{-3}\%$, respectively. As a result, the expected
total net gate fidelity was found: $\int d\delta\Phi_c
d\delta\epsilon\,
\mu(\delta\Phi_c)\mu(\delta\epsilon)F(\delta\Phi_c,\delta\epsilon)\sim99.99\%$.

The second measurement gate is in the conjugate basis. It evolves equal superpositions of $|0\rangle$ and $|1\rangle$ states
at the parking point,
$|\pm\rangle=\frac{1}{\sqrt{2}}(|0\rangle\pm|1\rangle)$, to the
final states $|L,0\rangle$ and $|R,0\rangle$ respectively, at the
measurement point. Since we know an adiabatic evolution would
preserve the amplitude of probability of each state in the
superposition, we have to design a non-adiabatic pulse in order to
implement this +/- measurement. Indeed, by passing at an
appropriate rate through the portal region, we achieve the desired
transformation. The second plot of Fig. \ref{measurement} presents
the proposed gate. It consists of two distinct parts: the first
occurs up to $t\lesssim7\rm{ns}$, when an adiabatic $\Phi_c$ pulse is
applied to the qubit. This part is done slowly to minimize the
leakage when the system passes through the avoided crossing gap;
it is also tuned so that the correct relative phase is accumulated
between the $|0\rangle$ and $|1\rangle$ states, with respect to a
reference phase. The second part of the pulse, $t\gtrsim\rm{7ns}$,
causes the qubit to undergo a non-adiabatic evolution through the
portal, producing the transitions needed to implement the gate.
The bias flux is maintained constant during the whole gate, and it
is held as close as is practical to the S line -- in our
calculation we take $\epsilon=30\mu\Phi_0$. The matrix representation Eq. (\ref{+-}) reveals the ideal transformation, which ends with an irrelevant relative phase between the ground and first excited states. For the path proposed,
the probability of leakage is found to be $0.03\%$.
Because the qubit has to stay, for an appreciable amount of time
in a region of very fast relative phase accumulation, the main
type of noise is a phase noise. The expected total net fidelity for
this gate is: $\int d\delta\Phi_c d\delta\epsilon\, \mu(\delta\Phi_c)\mu(\delta\epsilon)F(\delta\Phi_c,\delta\epsilon)\sim99.8\%$.

\subsection{Phase Gate}
The phase gate, Fig. \ref{zgate}, is the simplest operation of our
set. In order to accumulate the desired phase, {\it i. e.}
$\theta_z(\Delta t)\equiv\int_0^{\Delta t}\omega_{01}(t) dt$, it
is sufficient to adiabatically bring the system from the parking
point to a position where the $\omega_{01}$ frequency deviates few
hundreds MHz from the reference frame; then one just has to wait the
appropriate amount of time, corresponding to the desired phase
accumulation (see inset of Fig. \ref{zgate}). Since this position
can be reached without passing through the avoided crossing, the
leakage probability is extremely low; we calculate a leakage
probability  of $10^{-7}\%$. Since we remain essentially within a very stable parking regime, the
gate, for any value of $\theta_z$, is very insensitivity to
fluctuations of the both applied fluxes. Thus, the gate can be
performed with a very high total net fidelity:$\int d\delta\Phi_c d\delta\epsilon\, \mu(\delta\Phi_c)\mu(\delta\epsilon)F(\delta\Phi_c,\delta\epsilon)\sim99.999\%$. Since the
operation is designed to perform an adiabatic evolution of the
system, the 0/1 transitions are almost completely suppressed.
Consequently, the main source of (small) fidelity loss can be
characterized as a phase noise. Indeed, if we take a look at the operator-sum representation of the system superoperator, Eq. (\ref{operatorsum}), we obtain the following decomposition: $E_0\approx0.99999\hat{\sigma}_z$, $E_1\approx0.003 \hat{\mathds{1}}$, $E_2=0$ and $E_3=0$, which clearly indicates that we have just one form of noise.

\subsection{Hadamard Gate}

\begin{figure}[t!]
\begin{center}\includegraphics[ width=0.5\columnwidth,
 keepaspectratio]{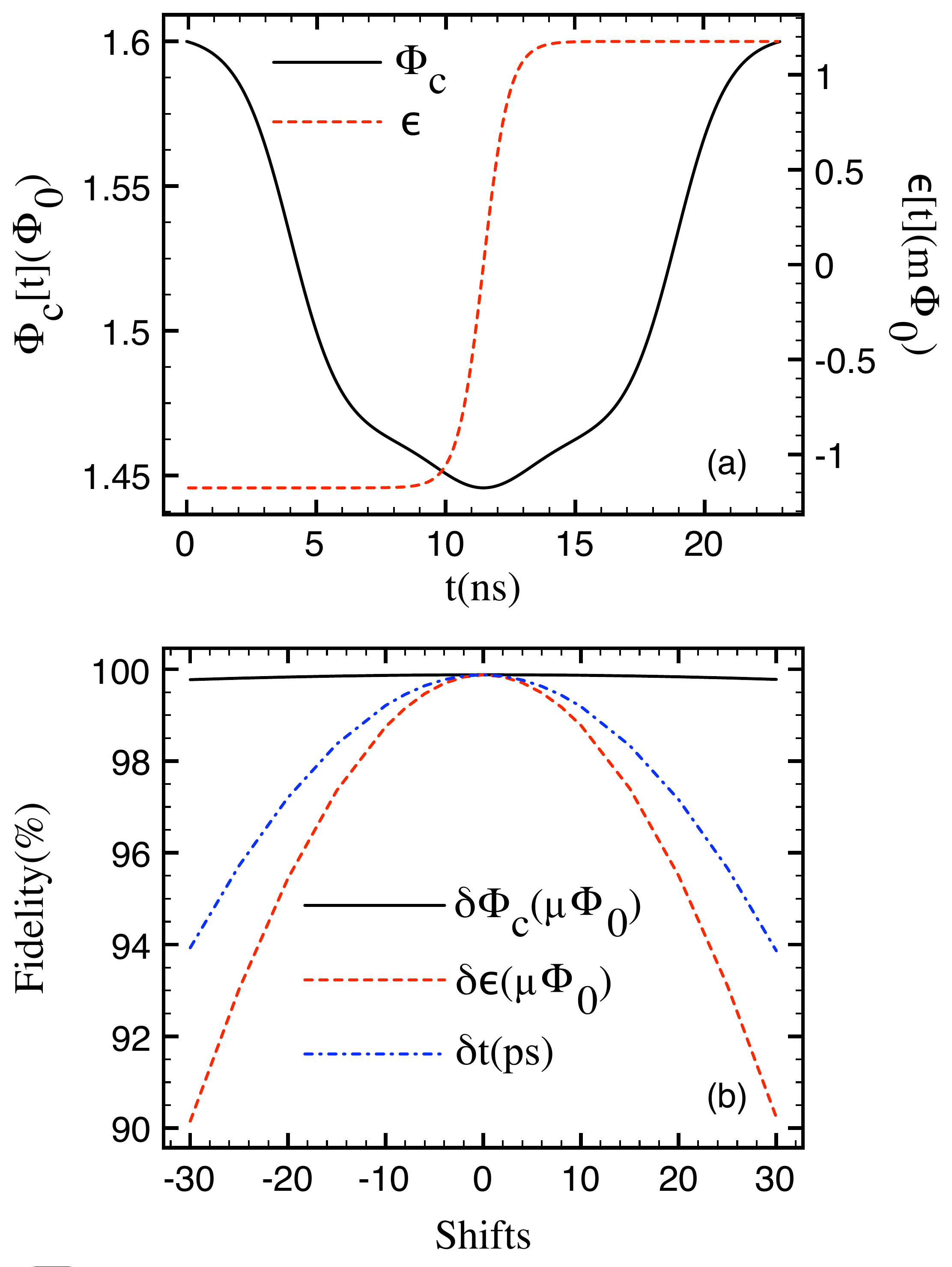}\end{center}
    \caption{The Hadamard gate. (a) The gate consists of a composition of pulses in $\Phi_c$ and $\epsilon$ fluxes. The $\Phi_c$ pulse is designed to bring the system from the memory regime to the portal in an adiabatic process. When the portal is reached, the system is kicked from one side of the S Line to the other through a non-adiabatic $\epsilon$ pulse, in order to obtain the wanted transitions. After applying the $\epsilon$ pulse, the system is brought adiabatically back to the memory position. (b) The fidelity of the gate as a function of unwanted shifts in $\Phi_c$, $\epsilon$, and the synchronization of the pulses, $\delta t$, from the optimal point of operation. The main source for the loss of fidelity is related with noise phase.}
    \label{hadamard}
\end{figure}

Completing our universal set of one-qubit gates is the Hadamard gate, Fig. \ref{hadamard}. Its implementation is more complex than the previous gates, since it requires the synchronization of the two parameters $\Phi_c$ and $\epsilon$. As one can see, the gate can be basically divided in three parts: the first, $t\lesssim10\rm{ns}$, evolves the qubit in an adiabatic process from the memory state to the portal. During this process, only control flux is changed; once reached the start of the portal, the second part of the gate is applied. It consists of a non-adiabatic pulse performed by the bias flux $\epsilon$. This pulse is responsible to very quickly move the qubit from on side of the S line to another. This process, $10\rm{ns}\lesssim t\lesssim 14\rm{ns}$, is designed to create the desired superpositions between states $|0\rangle$ and $|1\rangle$. In order to require reasonable rise-times for this pulse, we have to be as far as possible from the S line. Nevertheless, this excursion in the bias flux $\epsilon$ is limited due to the very small gap $\omega_{12}$ region seen in the flux space, see Fig. \ref{densitplots}; Finally, the last step, $t\gtrsim14\rm{ns}$, adiabatically brings the qubit back to the its initial position - again, only $\Phi_c$ is changed in this regime.

As highlighted in Fig. \ref{densitplots}, the path taken in the flux space due to this fluxes $\Phi_c(t)$ and $\epsilon(t)$ composition has the form of a ``U''-shape. This is a very useful feature, since working on both sides of the S line, one can expect to correct in the second half of the pulse some of the errors accumulated during the first. Indeed, because of the symmetry around the S line, in the first order of approximation, the errors due to the shifts in the bias flux, $\delta\epsilon$, should be canceled out, since at this order $\omega_{01}(\pm\epsilon+\delta\epsilon)=\omega_{01}(|\epsilon|)\pm\alpha\delta\epsilon$. However, for this to be true, a precise synchronization of the $\Phi_c(t)$ and $\epsilon(t)$ pulses is required, in order to obtain a symmetric path. For the paths in Fig. \ref{densitplots}, the number of dots indicates the rate as the system evolves: The fewer dots, the faster the time flux rate, one can clearly see the other strategies used to optimize the gate: at the parking regime, we can perform a very fast evolution, since the $\omega_{01}$ and $\omega_{12}$ are very big ($\sim 3.1\rm{GHz}$); once we have reached the avoided crossing position, in order to avoid leakage, we slow down the evolution. However, this regime also coincides with a fast phase accumulation rate, so we try to not spend too much time at this region. Thus, we find a trade-off that has to be worked out during the optimization of the gate. For the path shown, we expect to observe a probability of leakage of $0.12\%$.

The second plot of Fig. \ref{hadamard} shows the fidelity as a function of $\delta\Phi_c$, $\delta \epsilon$, and the flux desynchronization $\delta t$. It turns out the gate proposed is very insensitive to fluctuations in $\delta\Phi_c$. This occurs because we have designed the gate to explore a ``sweet'' spot in control flux. As one can see in Fig. \ref{spectrum},  the region close to $\Phi_c=1.447\Phi_0$ presents a first-order insensitive point to fluctuations of $\Phi_c$ for several values of $\epsilon$. Thus, since this point is at the portal, we have chosen it to ``sit'' the qubit while we perform the bias pulse. The average fidelity considering each noise channel separately is presented in Table \ref{table}. Because the $\delta\epsilon$ and $\delta t$ fluctuations essentially have the same effect, changing the relative phase of the two computational basis states, their loss of fidelity are found to be very similar. Thus, we see that the phase noise again plays the major role for the loss of fidelity. The complete noise characterization is obtained through the operator-sum representation of the system superoperator: $E_0\approx0.9953\hat{\rm{H}}$, $E_1\approx0.063 \hat{\sigma}_y$, $E_2\approx0.0048(i\hat{\mathds{1}}+(\hat{\sigma}_z-\hat{\sigma}_x)/4)$ and $E_3\approx0.0016(i\hat{\mathds{1}}/2+\hat{\sigma}_z-\hat{\sigma}_x)$. The net fidelity considering the effects of all noise channels together is given by: $\int d\delta\Phi_c d\delta\epsilon\,d\delta t \mu(\delta\Phi_c)\mu(\delta\epsilon)\mu(\delta t)F(\delta\Phi_c,\delta\epsilon,\delta t)\sim99.46\%$.

\section{\label{twoqubit}Two-qubit System}

The two-qubit system we have used to simulate the two-qubit gate is sketched in Fig.~\ref{qubit}. This layout preserves the same structures present for the one-qubit system, {\it i. e.} the transmission line, the readout SQUIDs and the flux lines; in addition, it has a qubit-qubit interaction that is assumed to arise due to the mutual inductance between the two big loops \cite{PhysRevB.60.15398} (other qubit-qubit coupling implementations using tunable interactions are demonstrated in \cite{hime,majer}). The qubit-qubit mutual inductance is considered to be a small parameter, such that a first-order perturbation theory is expected to give a very fair description of the system dynamics. Thus, following the procedure adopted to derive Eq. (\ref{4model}), we obtained the two-qubit system Hamiltonian given by (see Appendix)
\begin{eqnarray}
\fl\cH&=&H_A+H_B+H_I,\label{fullhamiltonian}\\
\fl H_I&=&\frac{1}{2}B'(\Phi_c^B)b(\Phi_c^A)\hat{\sigma}_z^A\otimes\hat{\mathds{1}}^B-\frac{1}{2}B'(\Phi_c^A)b(\Phi_c^B)\hat{\mathds{1}}^A\otimes\hat{\sigma}_z^B+J(\Phi_c^A,\Phi_c^B)\hat{\sigma}_z^A\otimes\hat{\sigma}_z^B.\label{qubit-qubit}
\end{eqnarray}
$H_{A,B}$ represent the single qubit Hamiltonians, Eq. (\ref{4model}), of the qubits A and B, respectively. The qubits are assumed to have identical bare qubits, but with different fundamental-mode transmission lines, $\omega_T^A\neq\omega_T^B$. This choice was made in order to avoid possible double excitation of the transmission lines due to the transfer of one quantum of energy from one transmission line to another. In addition, because we would like to use the same master clock to track the dynamics of both qubits, we had to impose a rational relation between $\omega_T^A$ and $\omega_T^B$. In our calculations, we have assumed that $\omega_T^A/\omega_T^B=3/4$.

\begin{figure}[t!]
\begin{center}\includegraphics[ width=0.5\columnwidth,
 keepaspectratio]{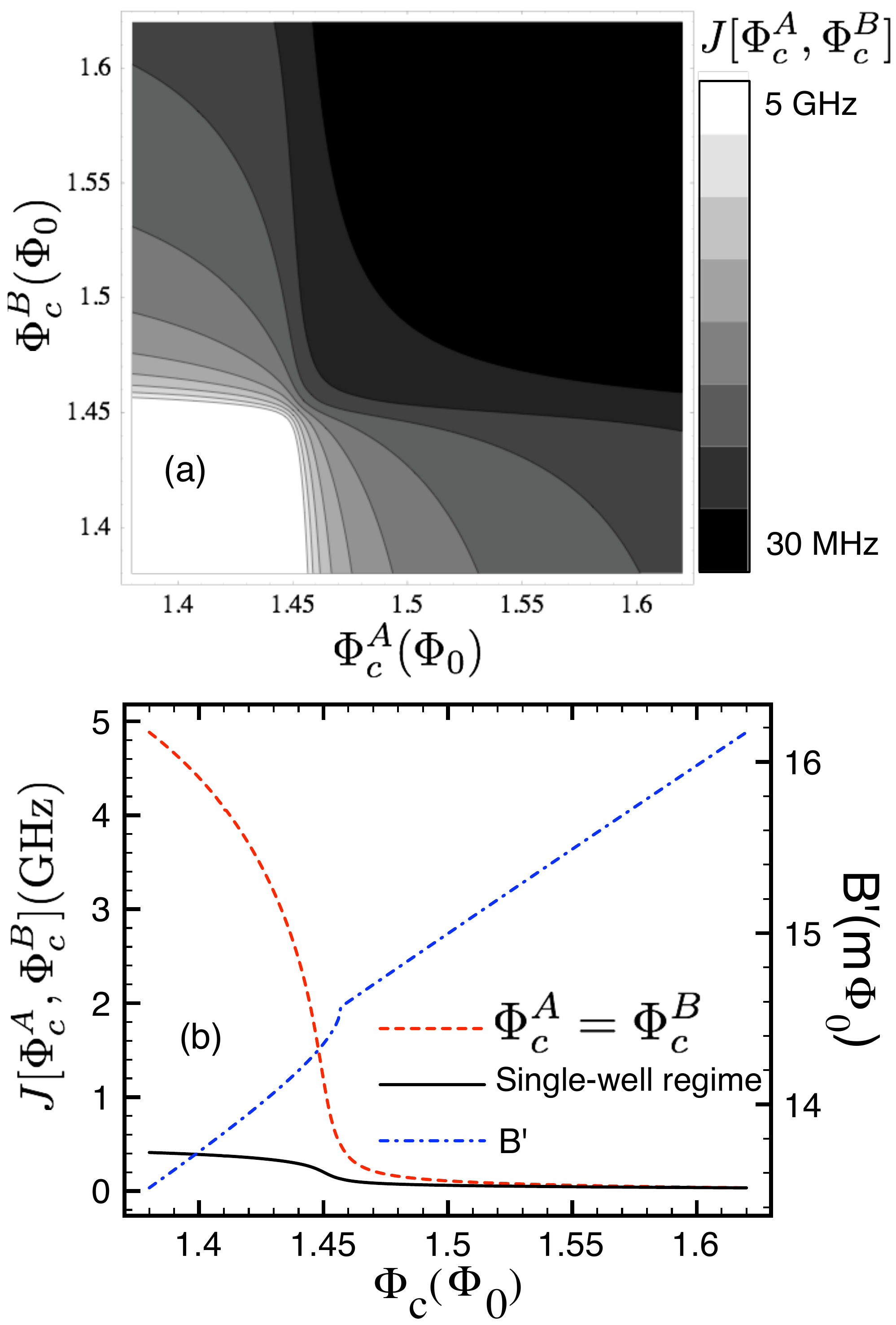}\end{center}
    \caption{(a) The effective qubit-qubit interaction $J$ as a function of the applied fluxes $\{\Phi_c^A,\Phi_c^B\}$. Even though the physical inductive interaction between the qubits is assumed constant, the effective qubit-qubit interaction is tunable as a function of the applied control flux in each qubit. This happens because of the change of the character of the system state when one changes the applied control flux in the qubit. (b) Solid curve presents $J$  when one of the qubits is maintained deeply in the single-well regime. Dashed curve shows the case when the qubits are moved in resonance through the control flux space. The bias flux $B'$ (dot-dashed curve) due to the circulating current of one of the qubits is presented as a function of its control flux.}
    \label{Jcoef}
\end{figure}

The system interaction Hamiltonian $H_I$ is given by Eq. (\ref{qubit-qubit}). Its terms may be understood as arising due to ``classical'' and ``quantum'' magnetic field components, in the following sense: For the first two terms, we clearly can identify them having a form of a bias field $B'$ applied to each qubit (compare with the bias term of Eq. (\ref{hamiltonianapp})). As expected, the bias field $B'$ applied to one qubit, let us say qubit $i$, depends of the control parameter of the other, qubit $j$. This happens because the external circulating current of qubit $j$ creates a magnetic field, $B'(\Phi_c^j)$, seen by qubit $i$. Nevertheless, for a given $\Phi_c^j$, the qubit $i$ feels the same bias field $B'(\Phi_c^j)$ whatever the quantum state of qubit $j$ is. Thus, $B'$ should be seen as a simply result of the Faraday's law applied to classical circuit: The field $\Phi_c^j$ induces a persistent current in the circuit $j$, which, in turn, creates a magnetic field $B'(\Phi_c^j)$ seen by the other qubit. Not surprisingly, the first two terms of $H_I$ cannot generate qubit-qubit entanglement. The dot-dashed curve in Fig. (\ref{Jcoef}) shows $B'$ a function of the control flux. Because of the presence of the Josephson junctions (non-linear inductances), $B'$ has non-linear behavior for some values of $\Phi_c$ (double-well regime).

As a consequence of the existence of the magnetic field $B'$, for the two-qubit system, the S line of qubit $i$ is  shifted to the lines $\epsilon^i=k\Phi_0\pm B'(\Phi_c^j)$ (where $k$ is an integer, and the choice of the sign depends on which loop the flux $B'$ is threaded through). Thus, we refer to those ``new'' qubit symmetric lines as the S' lines. This is a very important effect and must be taken into account when performing one-qubit operations for the two-qubit system.

The last term of Eq. (\ref{qubit-qubit}) describes a magnetic field seen by one qubit, determined by the quantum state of the other. The result is a qubit-qubit coupling of the form $\hat{\sigma}_z^A\otimes\hat{\sigma}_z^B$, which is responsible for generating qubit-qubit entanglement in the system. As one can see in Fig. \ref{Jcoef}, even though the physical qubit-qubit inductive coupling is assumed fixed, they effective interaction $J$ is tunable as a function of both control flux parameters. In fact, as shown in the second plot of Fig. \ref{Jcoef}, the coupling $J$ never reaches values higher than 450MHz when one qubit is maintained deeply in the single well regime (solid curve), rather than 5GHz observed when both qubits are at the measurement point (dashed curve). This is a direct manifestation of the change of the system state character as a function of control parameter.

In addition, because the transmission line of qubit A sees the qubit B only via interaction with its bare qubit, and the interaction between the bare qubits is in fact small, it turns out the transmission line states are very stable under the changes of control parameters of the other qubit. Indeed, since the qubit-qubit interaction Eq. (\ref{qubit-qubit}) is capable neither of moving the qubits from the single- to the double-well potential regime (for that terms of the form $\hat{\sigma}_x$ are needed) nor of changing the quantum number of the transmission lines, once one qubit is parked, the two-qubit system eigenstates stay ``frozen'' at the subspace in which the parked qubit eigenstates are close to the states $|S,0\rangle$ and $|S,1\rangle$. Thus, parking one qubit imposes a selection rule on the system that leads to a further reduction, below the already small value of $J$, of the effective qubit-qubit coupling. In fact, our calculations of the interaction matrix elements $|\langle i^A,j^B|J(\Phi_c^A,\Phi_c^B)\hat{\sigma}_z^A\otimes\hat{\sigma}_z^B|l^A,m^B\rangle|$ (where the states $i,j,l,m$ are the 0-1 one-qubit states) show they never reach values higher than 3MHz when one of the qubits is parked.

As a result, parking leads to a very effective way to decouple the qubits, thus allowing one to perform one-qubit gates for the two-qubit system discussed. By simply taking account of the fact that the qubit S lines in the flux space are moved to the S' lines, we can use exactly the same shaped pulse schemes presented previously for the one-qubit system, in order to perform the universal set of one-qubit gates. Indeed, our simulations showed that one can expect the same probability of leakage previously reported, and virtually the same expected fidelities, see Table \ref{table2}.

\begin{table}
\caption{\label{table2}The expected one-qubit gate fidelities, as a function of unwanted shifts in $\delta\Phi_c$, $\delta\epsilon$ and $\delta t$, when performed to the two-qubit system using the same schemes described in Section \ref{gates}. We observe a small difference between the one- and two qubit system fidelities only for the +/- measurement gate.}
\begin{indented}
\item[]\begin{tabular}{|c|c|c|c|c|}
\hline
\multicolumn{5}{|c|}{Two-qubit system}\\
\hline
Gate& & $\delta\Phi_c$ & $\delta\epsilon$ & $\delta t$\\
\hline
0/1 Measurement& $\langle\rm{Fidelity}\rangle(\%)$& $99.99$ & $99.99$ & -\\\hline
+/- Measurement & $\langle\rm{Fidelity}\rangle(\%)$& $99.76$ & $99.79$ &-\\\hline
Phase Gate & $\langle\rm{Fidelity}\rangle(\%)$& $99.999$ & $99.999$ &-\\ \hline
Hadamard & $\langle\rm{Fidelity}\rangle(\%)$& $99.87$ & $99.47$ & $99.63$\\\hline
\end{tabular}
\end{indented}
\end{table}

\subsection{Controlled-Z Gate}

\begin{figure}[t!]
\begin{center}\includegraphics[ width=0.5\columnwidth,
 keepaspectratio]{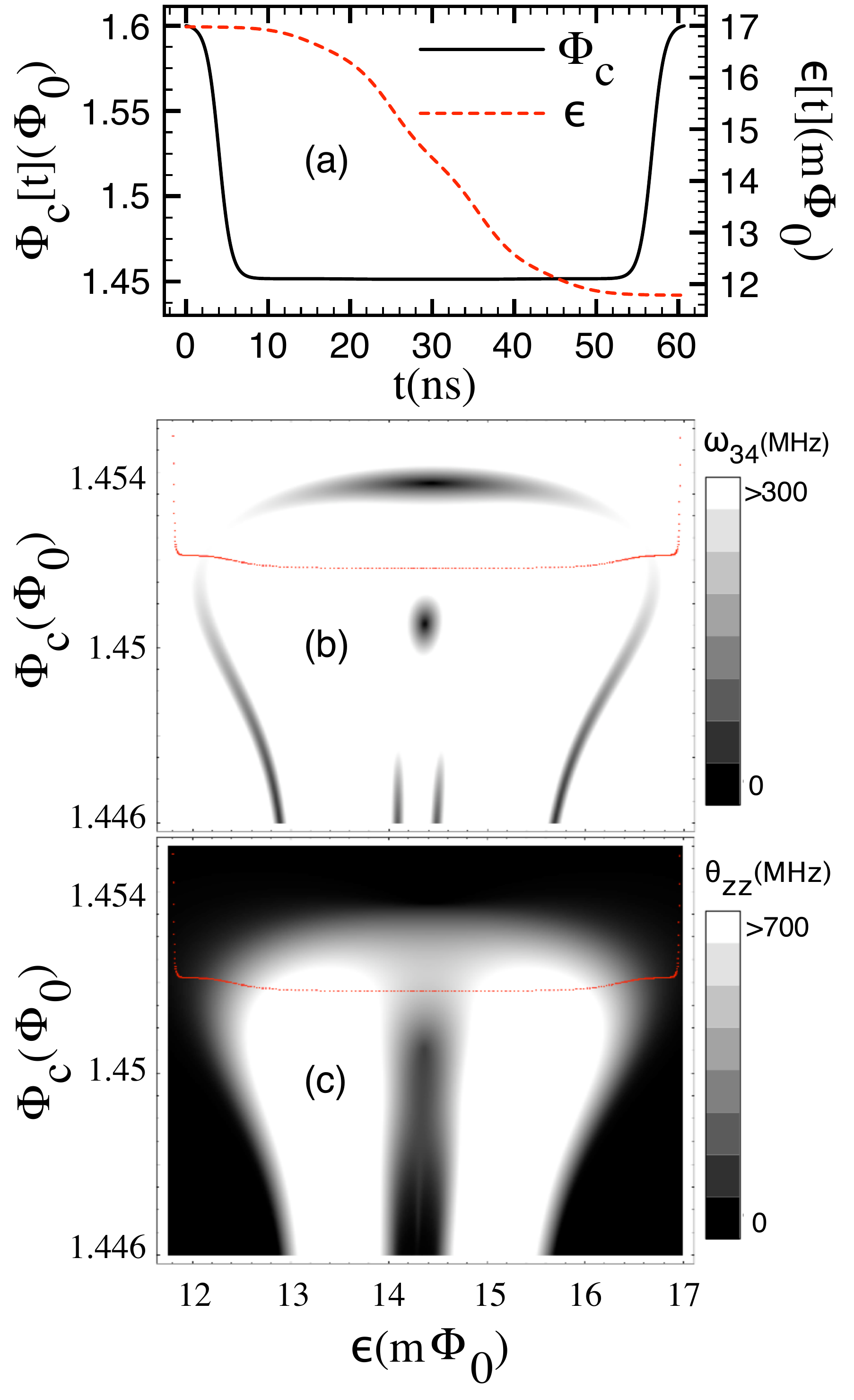}\end{center}
    \caption{(a) $\Phi_c(t)$ and $\epsilon(t)$ for the controlled-Z gate. The operation is performed adiabatically assuming $\Phi_c^A(t)=\Phi_c^B(t)$ and $\epsilon^A(t)=\epsilon^B(t)$. (b) The energy splitting $\omega_{34}$ and (c) the relative $\theta_{ZZ}$ for the flux space $\{\Phi_c^A=\Phi_c^B,\epsilon^A=\epsilon^B\}$. The energy splitting $\omega_{34}$ gives a limit on the rate at which the system can evolve without appreciable leakage. The $\theta_{ZZ}$ plot gives the phase accumulation rate of the principal parameter of the gate, Eq. (\ref{theta}). The path (a) taken in the flux space is illustrated with dots in (b) and (c). The number of dots indicates the time spent when passing through that region.}
    \label{figzz}
\end{figure}

The two-qubit gate proposed is a gate in the equivalence class\cite{PhysRevA.67.042313} of the controlled-Z gate, {\it i. e.} it only differs from the controlled-Z gate by local operations to the qubits A and B. The designed gate, first plot of Fig. \ref{figzz}, involves an adiabatic evolution of the system states, in order to accumulate the correct relative phases. Both $\Phi_c$ and $\epsilon$ of qubits A and B are used to perform the operation. In order to obtain the strongest qubit-qubit interaction, and thus the shortest gate possible, the control fluxes of both qubits are changed simultaneously, $\Phi_c^A(t)=\Phi_c^B(t)$. Also the bias fluxes change identically and, as for the Hadamard scheme, we work on both sides of the S' line. The desired unitary transformation has the matrix representation in the two-qubit eigenstate basis $\{|0\rangle,|1\rangle,|2\rangle,|3\rangle\}$ given by

\begin{equation}
U_{ZZ}=\left(\begin{array}{cccc}
e^{i\theta_{0}}&0&0&0\\
0& e^{i\theta_{1}} &0 &0\\
0&0&e^{i\theta_{2}}&0\\
0&0&0&e^{i\theta_{3}}
\end{array}\right).\label{zz}
\end{equation}

The relative phase
\begin{equation}
\theta_{ZZ}\equiv\theta_0-\theta_1-\theta_2+\theta_3\label{theta}
\end{equation}
is the pertinent parameter of this gate. Since the local invariants\cite{PhysRevA.67.042313} of the gate $U_{ZZ}$ are determined by
\begin{eqnarray}
G_1=\frac{1}{2}(1+\cos\theta_{ZZ}), \quad G_2=2(1+\frac{1}{2}\cos\theta_{ZZ}),
\end{eqnarray}
and those of the controlled-Z gate are given by $G_1=0$ and $G_2=1$, we end with the specific condition  $\theta_{ZZ}=\pi(1+2k)$ for the relative phase.

Since the evolution is done adiabatically, we can write the relative phase as $\theta_{ZZ}(\Delta t)=\int_0^{\Delta t} dt \{(E_0(t)+E_3(t))-(E_1(t)+E_2(t))\}$, where $E_i$ is the instantaneous eigenenergy of the state $i$. Thus, it is evident that $\theta_{ZZ}$ only has non-negligible values when the term $J\hat{\sigma}_z^A\otimes\hat{\sigma}_z^B$ of Eq. (\ref{qubit-qubit}) becomes appreciable. However, as shown in Fig. \ref{figzz}, to reach that region we have to pass through a regime of a very small gap ($\sim 300$MHz) between the $|3\rangle$ and $|4\rangle$ states (minimum gap between the computational basis $\{|0\rangle,|1\rangle,|2\rangle,|3\rangle\}$ and the rest of spectrum of the system). Because of that, our shaped dc gate has a total duration of 60ns in order to avoid leakage during the evolution. In addition, in order to explore ``sweet'' spots of operations, the designed $\Phi_c(t)$ pulse is not completely flat when passing through the minimum gap region (see Fig. \ref{figzz}(b) and (c)).

The gate is performed in the following way: starting from the memory state, both control fluxes are changed to the region around $\Phi_c\approx1.453\Phi_0$. Then, by changing the bias fluxes $\epsilon^{A,B}$, the qubits are slowly moved through the region of minimum gap $\omega_{34}$, passing from on side of the S' line to the other. At this stage, the parameter $\theta_{ZZ}$ becomes appreciable (hundreds of MHz). The gate is designed to spend the right amount of time, such that the final evolution will satisfy the necessary condition for $\theta_{ZZ}$. Once the other side of the S' line is reached, the qubits are brought back to the memory state. As observed for the Hadamard gate, the two-qubit gate proposed also has a ``U''-shape in the flux space.

The probability of leakage due to this process is $\sim0.04\%$.
Since the gaps between the computational states are bigger than
$700$MHz, unwanted transitions are strongly suppressed. Therefore,
phase noise is the main source of loss of fidelity.  As for
one-bit gates, this phase noise arises from physical 1/f noise and
pulse timing variations; we assume their magnitudes to be the same
as for the one-qubit case, with no correlations between the two
qubits.  We then calculate (assuming, again, Gaussian noise
statistics) the operator-sum representation of the system
superoperator:

\begin{eqnarray}
E_0\approx\sqrt{0.9905}(a_0\hat{\mathds{1}}^A\otimes\hat{\mathds{1}}^B+a_0^\ast\hat{\sigma}_z^A\otimes\hat{\sigma}_z^B\nonumber\\+c_0\hat{\mathds{1}}^A\otimes\hat{\sigma}_z^B-c_0^\ast\hat{\sigma}_z^A\otimes\hat{\mathds{1}}^B),\label{e0}
\end{eqnarray}

\begin{eqnarray}
E_1\approx\sqrt{0.0022}(a_1\hat{\mathds{1}}^A\otimes\hat{\mathds{1}}^B+i a_1^\ast\hat{\sigma}_z^A\otimes\hat{\sigma}_z^B\nonumber\\+c_1\hat{\mathds{1}}^A\otimes\hat{\sigma}_z^B-i c_1\hat{\sigma}_z^A\otimes\hat{\mathds{1}}^B),\label{e1}
\end{eqnarray}

\begin{eqnarray}
E_2\approx\sqrt{0.00014}(c_2\hat{\mathds{1}}^A\otimes\hat{\sigma}_z^B+i c_2\hat{\sigma}_z^A\otimes\hat{\mathds{1}}^B),\label{e2}
\end{eqnarray}
where $a_0\approx-(1+i)/2$, $c_0\approx-0.1-0.03i$, $a_1\approx0.4i-0.004$, $c_1\approx0.6-0.03i$ and $c_2\approx (1+i)/2$ \footnote{The ideal gate for the shaped pulse Fig. \ref{figzz} is such that $E_0\equiv\frac{1}{2}(1+i)\hat{\mathds{1}}^A\otimes\hat{\mathds{1}}^B+\frac{1}{2}(1-i)\hat{\sigma}_z^A\otimes\hat{\sigma}_z^B,~ {\rm and} ~E_k\equiv0~ {\rm for}~ k\neq0$.}. The other operation elements $E_k, ~k=3,\cdots, 16~$ are negligible. Thus, as one can see from Eqs. (\ref{e0}-\ref{e2}) these leading operation elements all lie at the subspace $\{\hat{\mathds{1}}^A,\hat{\sigma}_z^A\}\otimes\{\hat{\mathds{1}}^B,\hat{\sigma}_z^B\}$ revealing the phase character of the noise.

The total net fidelity considering the effects of all noise channels is given by
$\int d\delta\vec{\Phi}_c d\delta\vec{\epsilon}\,d\delta \vec{t} \mu(\delta\vec{\Phi}_c)\mu(\delta\vec{\epsilon})\mu(\delta \vec{t})F(\delta\vec{\Phi}_c,\delta\vec{\epsilon},\delta \vec{t})\sim99.67\%$

\section{\label{conclusion}Conclusion}

Obviously, it is hoped that the results of this paper will be of
direct relevance for experiments on qubits being performed at IBM,
as well as being of general relevance to the problem of precision
quantum control in any flux qubit system.  The noise mechanisms
analyzed here --- magnetic 1/f noise, Johnson noise of circuit
resistances, timing errors in pulse channels --- are the mixture
of fundamental and practical sources of error that are currently
understood in the laboratory.  Thus, the quantitative fact that
the values of the gate infidelity (one minus the fidelity) are at
the 1\% level and below, is the major result of this paper.

One might ask, given that this paper's main claim to importance is
quantitative, whether the approximations made in the quantum model
are actually adequate to give sub-1\% accuracy.  Perhaps the
fundamental circuit Hamiltonian may be accepted as being very
accurate; but can it be that the sequence of subsequent
approximations, namely 1) the representation of the transmission
line by a single oscillator mode, 2) the dimension-reduction
resulting from the Born-Oppenheimer approximation, and 3) the
final spectral truncation, involving an interpolated
oscillator-displacement parameter, resulting in our four-level
model, have the desired sub-1\% accuracy?  We can see, for
example, in Fig.~4, that the spectral truncation results in
changes in the absolute value of the energy eigenvalues, in some
places by as much as 7\%.

Despite this, we believe that our infidelity numbers are
nevertheless quite sound, so that if we calculate an infidelity
here of 1\%, it may be, in a more accurate calculation, actually
equal to 0.9\% or 1.1\%, but not much different from that.  We
have a few reasons for saying this.  First, for our gate designs,
it is generally not necessary that an energy gap be exactly a
certain value at a particular point on the flux axis; it is more
important that, somewhere in the near vicinity of a particular
flux value, the energy gap has a certain size.  This will be true
even for a model Hamiltonian of moderate accuracy.

Traced to a deeper level, the successful functioning of our gates
depends primarily on the following general features of the model:
1) Since many of the evolutions we consider are adiabatic, it is
important merely that some integrated properties of the energy
eigenvalues over some paths in parameter space be correct.  This
is fairly easily achieved by a model of moderate accuracy.  2) In
a few cases basis-changing, non-adiabatic evolutions are
important.  These are achieved in just three ways: by passing
through the portal, by crossing the S line, and (unintentionally)
by moving the system in and out of parking. Our model obviously
contains all of these features, with trends in the size of gaps
that certainly track those of an exact calculation very closely.
Furthermore, the general structure of the low-lying eigenvectors
are captured in our truncation, meaning that trends in the matrix
elements that determine the magnitude of Landau-Zener tunneling
effects, are also well represented.

So, we conclude that our qubit, as currently understood, should be
capable of gate operation at the 1\% noise level.  Unfortunately,
in the lab, in any given day (or month) there are ``bugs" in the
experiment that causes the qubit, often for unexplained reasons,
to have fidelities much worse than what we estimate here.  But we
nevertheless hope that these calculations will have significant
practical value.  Our gate set is universal, in one of two
different ways, in fact: the 0/1 measurement/preparation gates,
Hadamard, controlled-Z, and one-qubit phase gates form one well
known universal set \cite{nielsenchuang}; it is less well known that,
with more overhead, Hadamard can be replaced by the +/-
measurement/preparation gates \cite{kitaev-2006}.

So, can a ``debugged" IBM qubit be used soon for universal quantum
computation? The answer is, in our opinion, ultimately yes.  The
answer would certainly be no if the noise threshold for
fault-tolerant quantum computation were in the neighborhood of the
oft-quoted value of $10^{-5}$ \cite{gottesman,KLZ,aharonov}.  It
is not inconceivable for the experiment to get to these values
someday, since we find that the infidelities decrease much faster
than linearly with the assumed noise levels. (To get to $10^{-5}$
we would need to get to the very daunting levels of $100 {\rm
n}\Phi_0$ at 1Hz for the 1/f noise amplitudes and 100$fs$ for
timing accuracies; there is optimism that both of these numbers
are ultimately attainable.) Fortunately, while $10^{-5}$ was the
threshold as it was understood ten years ago, much recent work
shows that with good designs, much higher thresholds are possible
\cite{panosfibonacci,knill}.  According to \cite{knill}, $1\%$ is
in fact on the high end of the noise levels for which fault
tolerance may be possible.

But the final answer to the question of whether fault tolerance is
possible will require resolving a number of further uncertainties.
One effect not included here is the fact that noise will be
correlated between qubits that are physically nearby, due to
ordinary electrical crosstalk, say.  This is certainly deleterious
for noise thresholds.  However, a great virtue of our parking
scheme is that, while parked, qubits are quite immune to
electrical noise from nearby sources.  Another point that might
count in our favor is that, as seen in our operator-sum
expansions, the noise has a definite structure (i.e., it is mostly
phase noise in most cases), while most analyses of noise
thresholds have assumed worst-case, structureless noise.  On the
other hand, these special noise sources, to the extent that they
arise from the physical 1/f noise, are substantially correlated in
time. While fault tolerance is known to survive in the presence of
such ``non-Markovian" sources \cite{terhal-2005-71}, it is not clear
what such an effect does to the numerical values of the threshold.
Finally, there are purely ``architectural" considerations, e.g.,
the theoretical analyses assumed that a qubit can be moved to the
proximity of any other one without cost.  This will surely not be
true in any real Josephson architecture.

To summarize, we have demonstrated that using only low-bandwidth
electrical pulses, a universal set of high fidelity quantum gates
can be achieved for an oscillator-stabilized Josephson qubit, with
a required ``clock time" around 60ns, and achievable fidelities
above 99\% per gate operation.  Essential to these results are the
existence of ``parking" made possible by coupling of the flux
qubit to a transmission line, the possibility of mostly adiabatic
control, and the availability of two robust basis-changing effects
obtained by crossing the ``portal" and the ``S line".  Time will
tell if these physical elements make it possible, in the face of
the many sources of noise that are present in the solid state
environment, to do large-scale quantum computation.

\appendix

\section{Hamiltonian derivation}

In this appendix we supply derivations of the Hamiltonians presented in the main text, Eq. (\ref{hamiltonianapp}) and Eq. (\ref{qubit-qubit}). The main assumption in the derivation is the use of first-order perturbation theory to treat terms of the system potential. As presented in Fig. \ref{4model} and previously discussed, performing the procedure described here, we could arrive at a very good description for the lowest states of the IBM qubit.

\subsection{Brief review of BKD}

Burkard, Koch and DiVincenzo \cite{Burkard:2004aa} have introduced a universal method for analyzing any electrical circuit containing Josephson junctions, provided only that its elements can be represented by lumped elements. The methodology can be summarized in a few steps: A network graph is written for the circuit. A network graph is simply a drawing of the circuit where each two-terminal element (inductor, capacitor, Josephson junction, etc.) is represented as an oriented labeled branch connecting two nodes. Then, a {\it tree} of the network graph (a subgraph that does not contain any loops) is chosen using the following criteria: the tree has to contain all of the capacitors in the system, no resistors or external impedances, no current sources, no Josephson junctions and as few linear inductors as possible. The branches do not belong to the tree are called {\it chords}.

Associated with the tree chosen, there are the so-called sub-loop matrices $\bF_{XY}$, where the label $X$ represents the tree elements ($X=C$ for capacitors and $X=K$ for the tree inductors) and $Y$ the chord branches ($Y=J$ for Josephson junctions, $Y=L$ for linear inductors, $Y=R$ for shunt resistors, $Y=Z$ for external impedances and $Y=B$ for bias current sources). The sub-loop matrices have entries -1, 0, or 1, and give information about the interconnections in the circuit, determining which tree branches $X$ are present in which loop defined by the chords $Y$. The entries in $\bF_{XY}$ matrix are found as follows: If the corresponding tree element $X_i$ does not belongs to the loop defined by the corresponding chord $Y_j$, its entry $F_{XY}^{i,j}$ is zero. If it belongs to the loop but has opposite orientation to $Y_i$, we have $F_{XY}^{i,j}=+1$. Finally, if it belongs and has the same orientation, we have $F_{XY}^{i,j}=-1$. These steps give an algorithm to encode the topology of the system in a matrix representation. In addition, the formalism assumes that all capacitors should be considered to be in parallel with a Josephson junction, even if it is one with zero critical current.

The physics of the circuit is introduced by imposing the Kirchhoff laws at each node of the network, and defining the electrical characteristics of each branch type as
\begin{eqnarray}
{\bf I}_J&=&{\bf I}_{\rm c}\sin{\boldsymbol \varphi},\label{junction}\\
{\bf Q}_C&=&{\bf C} {\bf V}_C,\label{capacitor}\\
{\bf V}_R&=&{\bf R}{\bf I}_R,\label{resistor}\\
{\bf V}_Z(\omega)&=&{\bf Z}(\omega){\bf I}_Z(\omega),\label{impedance}\\
 \left(\begin{array}{c}{\boldsymbol \Phi}_L\\ {\boldsymbol
   \Phi}_K\end{array}\right)&
  = & \left(\begin{array}{l l}{\bf L}        & {\bf L}_{LK}\\
                           {\bf L}_{LK}^T & {\bf L}_{K} \end{array}\right)
   \left(\begin{array}{c}{\bf I}_L\\ {\bf I}_K\end{array}\right).\label{inductance}
\end{eqnarray}
Here the diagonal matrix ${\bf I}_{\rm c}$ contains the critical currents of the junctions, and $\sin\boldsymbol \varphi$ is the vector $(\sin\varphi_1,\sin\varphi_2,\cdots,\sin\varphi_{N_J})$. The phase $\varphi_i$ represents the superconducting phase difference across the junction $i$. The linear capacitors are described by Eq. (\ref{capacitor}), with their capacitance values given by the entries of the diagonal matrix $\bC$. The junction resistors are assumed to follow Ohm's law, Eq. (\ref{resistor}), where $\bf R$ is the real and diagonal shunt resistance matrix. The impedances are described by Eq. (\ref{impedance}), that relates the Fourier transforms of the current and voltage. ${\bf Z}(\omega)$ is the diagonal impedance matrix. Since linear inductors can be tree branches as well as chords, we have to distinguish them to apply the network graph theory. Therefore, the inductance matrix must be organized in the block form shown in Eq. (\ref{inductance}), where $\bL$ is the inductance matrix of the tree inductors, $\bL_K$ is that for the chord inductors, and $\bL_{LK}$ represents the mutual inductances between the tree and chord inductors.

BKD arrived at the system Hamiltonian
\begin{eqnarray}
\fl \cH_S(t)&=&\frac{1}{2}\bQ_C^T\bC^{-1}\bQ_C+\left(\frac{\Phi_0}{2\pi}\right)^2 U(\boldphi,t),\label{BKDhamiltonianappendix}\\
\fl U(\boldphi,t)&=&-\sum_iL_{J;i}^{-1}\cos\varphi_i+\frac{1}{2}\boldphi^T\bM_0\boldphi
+\frac{2\pi}{\Phi_0}\boldphi^T[(\bar{\bN}*\tilde{\boldPhi}_x)(t)+(\bar{\bS}*\bI_B)(t)],\label{BKDpotentialappendix}
\end{eqnarray}
where the external applied fluxes and current are respectively represented by  $\tilde{\boldPhi}_x$ and $\bI_B$. The first term of Eq. (\ref{BKDpotentialappendix}) is due to the Josephson energies, where $L_{J;i}^{-1}\equiv\left(\frac{2\pi}{\Phi_0}\right)I_{c;i}$ and $I_{c;i}$ is the critical current of junction $i$. The  second is associated with the inductive energies of each branch of the circuit. The topology of the circuit is encoded in the matrices matrices $\bM_0$, $\bar{\bN}$ and $\bar{\bS}$, given by Eqs. (62), (63) and (66) in \cite{Burkard:2004aa}. Thus, BKD maps the circuit dynamics to that of a massive particle in a potential, whose masses and degrees of freedom are associated with the system capacitances. The quantization of the system is introduced by imposing the canonical commutation relation for the variables of charge, $\bQ_c$, and phase $\boldphi$: $[\frac{\Phi_0}{2\pi}\varphi_i,Q_{C;j}]=i\hbar\delta_{ij}$.
\subsection{IBM qubit network graph}
\begin{figure}[t!]
\begin{center}\includegraphics[ width=0.5\columnwidth,
 keepaspectratio]{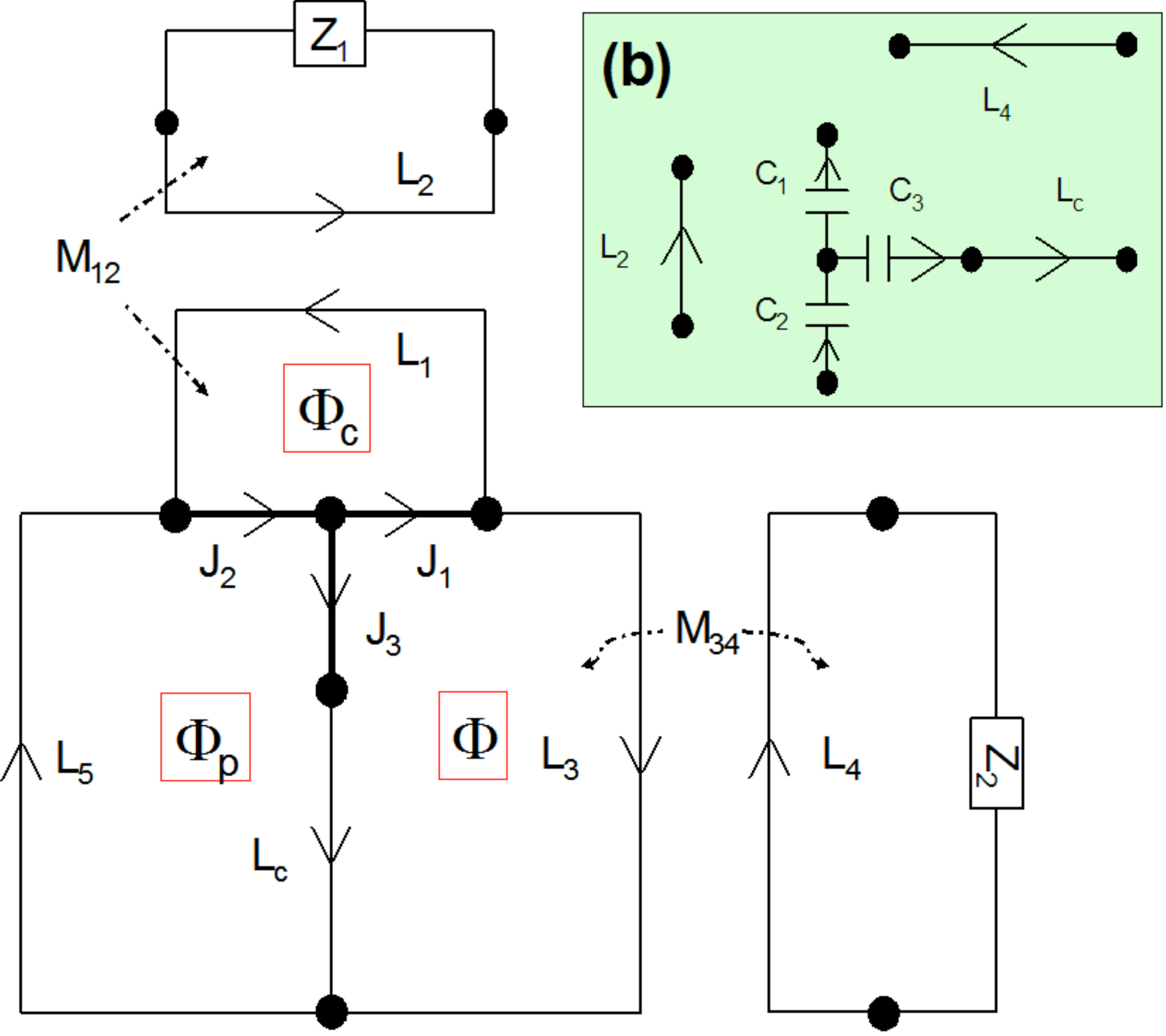}\end{center}
    \caption{The network graph for the bare qubit. Each Josephson junction $J_i$ (thick line) is modeled by a resistively shunted Josephson junction (RSJ) containing an ideal junction with critical current $I_{c,i}$, a parallel junction capacitance $C_i$ and a shunt resistance $R_i$. The qubit is coupled to the flux lines via mutual inductances $M_{12}$ and $M_{34}$. Because of the ``gradiometer'' structure of the qubit, the ``pick up'' flux line (responsible to apply the flux $\Phi_{\rm p}$) was omitted without loss of generality. Inset (b): tree chosen for the graph.}
    \label{tree}
\end{figure}

Fig. \ref{tree} shows the network graph and the chosen tree for the bare qubit. The sub-loop matrices can be read off by inspection:
\begin{equation}
F_{CL}=\left(\begin{array}{rrr}-1&-1&0\\-1&0&-1\\
0&1&-1\end{array}\right),\,\,
F_{CZ}=\left(\begin{array}{rr}0&0\\0&0\\
0&0\end{array}\right),
\end{equation}
\begin{equation}
F_{KL}=\left(\begin{array}{rrr}0&0&0\\0&0&0\\
0&1&-1\end{array}\right),\,\,
F_{KZ}=\left(\begin{array}{rr}-1&0\\0&-1\\
0&0\end{array}\right).
\end{equation}

The {\bf L} matrices are denoted:
\begin{equation}
L=\left(
\begin{array}{lll}
L_1 & M_{13} & M_{15} \\
 M_{13} & L_{3} & M_{35}  \\
 M_{15} & M_{35} & L_{5}  \\
\end{array}
\right),\end{equation}
\begin{equation}
L_{LK}=\left(
\begin{array}{lll}
 M_{12} & 0 & M_{1c} \\
 0 & M_{34} & M_{3c} \\
 0 & 0 & M_{5c} \\
 \end{array}
\right),
\end{equation}

\begin{equation}
L_{K}=\left(
\begin{array}{lll}
 L_2 & 0 & 0 \\
 0 & L_4 & 0 \\
 0 & 0 & L_{c}
\end{array}
\right).
\end{equation}

Using the modeling program FASTHENRY, we obtained the following inductance values for our qubit: $L_1=32.18,~L_3=L_5=686.93,~L_2=32.18,~L_4=605.03,~L_c=106.27,~M_{13}=M_{15}=-2.23,~M_{35}=-12.25,~M_{12}=0.8,~M_{34}=0.5,~M_{1c}=0({\rm exact}),~M_{3c}=-M_{5c}=28.15$ (all in units of pH).

In addition, the capacitances and critical currents of the junctions are taken to be $C=10$fF and $I_c=1.3\mu$A.

\subsection{Bare qubit Hamiltonian}

Our aim in this section is to present the steps used to derive the approximate Hamiltonian for the bare qubit. In order to reach the appropriate conditions to perform a first-order perturbation theory to treat the coupling between the qubit and the bias flux line, we firstly have to identify the ``slow'' and ``fast'' coordinates of the system. The slow coordinate is that that joins the two minima when the system presents a double-well structure, or that which has the slowest curvature in the single-well regime. The fast coordinates are those in which the potential rises very steeply, such that the system dynamics is frozen into the ground state along these directions.

The procedure to identify the slow and fast degrees of freedom starts from the exact system potential Eq. (\ref{BKDpotential}) and a unitary transformation $\bf R$ that diagonalizes $\bM_0$ ($\bM_0$ is a real and symmetric matrix \cite{Burkard:2004aa}). Using $\bf R$ we can decouple the quadratic form of the potential by transforming the system coordinate to the new coordinates ${\bf n}\equiv{\bf R}\boldphi$. Thus, from Eq. (\ref{BKDpotential}) we obtain (ignoring the term due to external sources of current)
\begin{eqnarray}
U({\bf n},t)=&-&\sum_iL_{J;i}^{-1}\cos\left(\sum_k r_{ik}n_k\right)\nonumber\\
&+&\frac{1}{2}{\bf n}^T{\boldsymbol \lambda}{\bf n}+\left(\frac{2\pi}{\Phi_0}\right){\bf n}^T{\boldsymbol \Phi}_x,\label{rotatedpotential}
\end{eqnarray}
where we have defined the diagonal matrix ${\boldsymbol \lambda}\equiv{\bf R}\bM_0{\bf R}^T$, which entries are the exact eigenvalues $\lambda_1=0, \lambda_2=3/(L_3+2L_c-4M_{3c}-M_{35})~ {\rm{and}}~ \lambda_3=(L_1+2(L_3+M_{35}-2M_{15}))/(L_1(L_3+M_{35})-2M_{15}^2)$, and the ``non-orthogonal'' flux vector ${\boldsymbol \Phi}_x\equiv{\bf R}\bar{\bN}\tilde{\boldPhi}_x=\{0,\frac{\lambda_2}{\sqrt{6}}\epsilon,-\frac{\sqrt{2}(L_3-M_{15}+M_{35})}{L_1(L_3+M_{35})-2M_{15}^2}\Phi_c\}$, with
\begin{eqnarray}
\epsilon&\equiv&\Phi-\Phi_{\rm p},\label{epsilon}\\
\Phi_c&\equiv&\tilde{\Phi}_c+\frac{L_1-2M_{15}}{2(L_3+M_{35}-M_{15})}(\Phi+\Phi_{\rm p}).\label{newphi}
\end{eqnarray}
Because of $\frac{L_1-2M_{15}}{2(L_3+M_{35}-M_{15})}\approx0.024$ and the fact we expect to work with magnetic fluxes of the order of only few $\Phi_0$s, we have that, to good approximation, the system potential is only a function of the difference of the magnetic flux in the two large loops, defined as $\epsilon$. This property is a direct manifestation of fact the designed bare qubit has a ``gradiometer'' structure. Even though this correction is expected be very small, we shall consider it during the whole procedure we are going to describe here.

The new coordinate system allows us to easily identify, by inspection of each term of Eq. (\ref{rotatedpotential}), the following symmetry in the potential
\begin{eqnarray}
U\left(\textbf{n},\tilde{\Phi}_c,\Phi,\Phi_{\rm p}\right)&=& U\left(n_1,n_2,n_3,\Phi_c,\epsilon\right)\nonumber\label{potentialsymmetry}\\
&=&U\left(-n_1,-n_2,n_3,\Phi_c,-\epsilon\right).
\end{eqnarray}
This shows that the system potential presents a definite symmetry in the plane determined by the directions $\{n_1,n_2\}$, when the system goes from $+\epsilon$ to $-\epsilon$ (if we simultaneously adjust the control flux $\tilde{\Phi}_c$ to maintain the new flux coordinate $\Phi_c$, Eq. (\ref{newphi}), constant). In addition, Eq. (\ref{potentialsymmetry}) also permits us to determine the conditions for which the system presents the same physical potential
\begin{eqnarray}
U\left(n_1,n_2,n_3,\Phi_c,\epsilon\right)=U\left(n_1',n_2',n_3',\Phi_c',\epsilon'\right) +\gamma,\label{slines}
\end{eqnarray}
where $\gamma$ is a irrelevant constant. In fact, we find that whenever the bias flux difference $\Delta\epsilon=\epsilon'-\epsilon$ is an integer multiple of the flux quantum, let us say $\Delta\epsilon=(k_1-k_2-2k_3)\Phi_0$ (each $k_i$ is any integer), the Eq. (\ref{slines}) is satisfied if we have the flux shift $\Delta\Phi_c=\Phi_c'-\Phi_c=\frac{L_1+2(L_3+M_{35}-2M_{15})}{2(L_3+M_{35}-M_{15})}\Phi_0(k_1+k_2)$ accompanied by the Josephson phase changes $\Delta\varphi_1=2\pi k_1,~\Delta\varphi_2=2\pi k_2,~{\rm and}~\Delta\varphi_3=2\pi k_3$.

It is now worth introducing a shifted coordinate system defined by ${\bf n}={\bf s}+{\bf d}$, where ${\bf d}\equiv \{0,0,\frac{1}{\lambda_3}\left(\frac{2\pi}{\Phi_0}\right) \frac{\sqrt{2}(L_3-M_{15}+M_{35})}{L_1(L_3+M_{35})-2M_{15}^2}\Phi_c\}$. The potential written in this coordinate system is given by
\begin{eqnarray}
\lefteqn{U({\bf s},\Phi_c,\epsilon)=\frac{1}{2}(\lambda_2 s_2^2+\lambda_3 s_3^2)+\left(\frac{2\pi}{\Phi_0}\right)\frac{\lambda_2}{\sqrt{6}}\epsilon s_2\nonumber}\\
&{}&-\sum_iL_{J;i}^{-1}\cos\left(\sum_k r_{ik}s_k+\frac{r_{i,3}}{\lambda_3}\left(\frac{2\pi}{\Phi_0}\right) \frac{\sqrt{2}}{L_1}\Phi_c\right).\qquad\quad\label{shift2}
\end{eqnarray}
Observe that the shifted coordinate system $\bf s$ is such that we still have the symmetry observed in Eq. (\ref{potentialsymmetry}): $U(s_1,s_2,s_3,\Phi_c,\epsilon)=U(-s_1,-s_2,s_3,\Phi_c,-\epsilon)$. However, since $\lambda_2/\lambda_3\approx0.059$ and $L_{j;i}^{-1}/\lambda_3\approx0.061$ ({\it i. e.} the potential is much steeper in the direction $s_3$ than in the others), we have the global minimum of the potential occurring at the position $s_3\approx0$. As a result, the bare qubit low-level dynamics can be considered frozen at the plane $s_3\approx0$.

Another important result is derived from the fact the low-level system dynamics is only governed by the degrees of freedom $s_1$ and $s_2$: the system potential has a symmetry line (S line) at $\epsilon=0$ in the flux space. Indeed, as one can see, when $\epsilon=0$, $U({\bf s},\Phi_c,\epsilon)$ has a perfect symmetry around the origin in the $\{s_1,s_2\}$ plane. Because of that, the low-level system wave functions have definite parity symmetry at $\epsilon=0$. In addition, because the same physical potential is found when $\epsilon$ is changed by an integer multiple of $\Phi_0$, a S line occurs whenever $\epsilon$ is an integer multiple of $\Phi_0$.

Unfortunately, because $\lambda_2\sim L_{j;i}^{-1}$, it is not simple, as it was for the direction $s_3$, to determine the soft and fast directions in the plane defined by $s_3\approx0$. However, from the symmetry of the S line one can see that, if the potential has a minimum in $s_1^{min}\neq0$ and $s_2^{min}\neq0$, the potential presents a symmetric double-well structure (with maximum at $s_1=s_2=0$) in the direction connecting the minimum points. This direction is expected to be the slow coordinate of the system, while its perpendicular direction should determine the fast degree of freedom. Consequently, by finding the minima positions in the plane $\{s_1,s_2\}$, one can determine those directions. Following this procedure, we perform one more rotation, $q_1=s_1\cos\theta-s_2\sin\theta $, $q_2=s_1\sin\theta +s_2\cos\theta$ and $q_3=s_3$, such that the direction $q_1$ connects the minima. When the potential does not have a double-well structure, we define the last rotation so that the potential curvature in the direction $q_1$ is the smallest.

Thus, after appropriate transformations, we end with phase and flux coordinate systems in which the symmetries presented by the system are much more clearly stated. In addition, the slow, $q_1$ and fast, $\{q_2,q_3\}$, degrees of freedom of the system are also identified. As already mentioned, because the potential is very steep the fast directions, the low-level system dynamics is frozen into the ground state along these direction. Following the Born-Oppenheimer approach developed in \cite{DiVincenzo:2006aa}, those directions can be traced out and their effects incorporated as small corrections to the remaining slow-coordinate potential energy.

At last, following all the steps described above, we are now in position to construct term by term of Hamiltonian Eq. (\ref{4model}):
\begin{itemize}
\item {\it No bias flux line} ($\epsilon=0$): As the system potential has a perfect symmetry around the origin, the bare qubit wave functions also have a parity symmetry. Thus, the ground and first excited states are expected to be symmetric and antisymmetric with respect to reflection around the origin. As described in the main text, the system potential is found in a double-well structure for small values of $\Phi_c$. There the bare qubit eigenstates can be understood as symmetrical and antisymmetrical equal superpositions of the classical orbital states of the left and right wells. For the single-well regime, $\Phi_c\gtrsim1.5\Phi_0$, the bare qubit states are associated with those of a harmonic oscillator. Thus, in this representation, the Hamiltonian of the bare qubit without bias flux line can be written as $-\frac{1}{2}\Delta(\Phi_c)\hat{\sigma}_x$, where $\Delta(\Phi_c)$ (shown in Fig. \ref{hamiltonian}) is the ground and first excited state energy splitting.
\item {\it The bias flux term}: As previously discussed, the bias flux $\epsilon$ is responsible for breaking the perfect potential symmetry observed at the S line. However, the system presents, as stated in Eq. (\ref{potentialsymmetry}), another important symmetry in going from one side of the S line to another ({\it i. e.} passing from $-\epsilon$ to $+\epsilon$): the system potential at $\epsilon<0$ is identical to that in $\epsilon>0$ by a mirror reflection (a $\pi$ rotation for the full multi-dimensional potential) at the origin. In the double-well regime that corresponds to the interchange of the left and right states under the mentioned transformation. The bias term of the Hamiltonian Eq. (\ref{4model}) is obtained applying a first-order perturbation theory to treat the bare qubit-bias flux line coupling. Indeed, because the potential term representing their coupling, $U_{q\epsilon}\approx\epsilon q_1 \pi\sqrt{\frac{2}{3}}\lambda_2 \sin\theta$, is very small, for typical values of bias flux used in this work ($\epsilon=O({\rm m}\Phi_0)$), compared to the other potential terms, the first-order approximation can be done in a very controllable way. Because of the symmetry of the bare qubit eigenstates, the interaction matrix elements $\langle \Psi_i |U_{q\epsilon}|\Psi_j\rangle$ (where $\Psi=S, A$ are the symmetrical and antisymmetrical bare qubit eigenstates) are non-zero only for those connecting the ground and excite states. Thus, the bias term has the form: $\frac{1}{2}b(\Phi_c)\hat{\sigma}_z$, with
\begin{equation}
b(\Phi_c)\equiv 2\pi\sqrt{\frac{2}{3}}\lambda_2 \sin\theta \langle S|q_1|A\rangle.
\label{bequation}
\end{equation}
\end{itemize}

\subsection{Bare qubit-transmission line coupling}

Another structure present in the IBM qubit is a high-quality superconducting transmission line Fig. \ref{qubit}. We model the fundamental mode of the transmission line as a simple LC circuit of definite frequency $\omega_T$ coupled to the bare qubit. We assume its characteristic impedance and inductance are given by $Z_0=\sqrt{\frac{L_T}{C_T}}=110\Omega$ and $L_T=5.6$nH respectively, and the bare qubit-transmission line mutual inductance equals $M_{qT}=200$pH. In order to find the slow and fast coordinates, we apply the same steps discussed above. An interesting feature of the new system is the switch of the slowest coordinate from the bare qubit degree of freedom to the transmission line coordinate. Indeed, we observe for small values of $\Phi_c$, when the bare qubit ground and first excited state energy splitting is much less than $\hbar \omega_T$, that the slow coordinate is associated with that of the bare qubit. However, when the energy splitting is higher than $\hbar\omega_T$, at high values of $\Phi_c$, the slowest coordinate is that associated with the transmission line degree of freedom. Because of that, when applying the Born-Oppenheimer approximation, for a correct description, we stop with a two dimensional potential energy of coordinates $q_1$ for the bare qubit and $q_T$ for the transmission line.

Provided that the bare qubit-transmission line coupling $U_{qT}$ is very small compared to the other potential terms: $U_{qT}/\lambda_2\approx 10^{-3}$, we can safely apply a first-order perturbation theory to treat that coupling. Since the interaction term has the form $U_{qT}\propto q_1 q_T$, the only non-zero interaction matrix elements $\langle \Psi_i^q\varphi_m^T |U_{qT}|\Psi_j^q\varphi_n^T\rangle$ (where $\Psi=S, A$ are the symmetrical and antisymmetrical bare qubit eigenstates, and $\varphi=0,1,\cdots,n$ are the harmonic oscillator states) are those connecting the ground and first excited bare qubit states and the transmission line states differing by one quantum of energy. Thereby, we obtain the Hamiltonian form: $g(\Phi_c)(\hat{a}+\hat{a}^\dagger)\hat{\sigma}_z$, with
\begin{equation}
g(\Phi_c)\propto (\langle S |q_1|A\rangle) (\langle 0|q_T|1\rangle)
\label{gequation}
\end{equation}

\subsection{Qubit-qubit coupling}

The layout of the two-qubit system is shown in Fig. \ref{qubit}. This scheme contains the same structures as the one-qubit system: the transmission line, the readout SQUIDs and the flux lines. The bare qubits are assumed identical, but, in order to avoid possible double excitations of the transmission line states due to the transference of quanta of energy from one transmission line to another, the transmission line fundamental mode frequencies are made different $\omega_T^A\neq\omega_T^B$. As highlighted in Fig. \ref{qubit}, the qubit-qubit interaction arises due to the mutual interaction between the two big loops, which is assumed to be by $M=12$pH.

Following the steps introduced for the bare qubit Hamiltonian, we reach after the diagonalization of the $\bM_0$ matrix the system potential
\begin{eqnarray}
{\cal U}({\bf n}^A,{\bf n}^B,t)=U({\bf n}^A,t)+U({\bf n}^B,t)+U^{AB}
\end{eqnarray}
where $U({\bf n}^i,t)$ corresponds to the potential Eq. (\ref{rotatedpotential}), and $U^{AB}=\sum_{ij}c_{ij}n_i^A n_j^B$ is the interaction potential. We find that $c_{ij}^{max}/\lambda_2\approx 10^{-3}$, what justifies considering the qubit-qubit coupling a small parameter of the potential, and consequently that the qubit dynamics is weakly perturbed by their interaction. Thus, we can perform the same steps described for the one-qubit system to find the qubits slow and fast coordinates.

An interesting result arises when the qubit potentials are written in the shifted coordinate system Eq. (\ref{shift2}). Under this transformation, the interaction potential $U^{AB}$ changes to the form
\begin{equation}
U^{AB}=\sum_{i,j}c_{ij}s_i^A s_j^B+\sum_{i}( c_{i3}d_3^B s_i^A+ c_{3i}d_3^A s_i^B),\label{qqinteraction}
\end{equation}
where $d_3^i\equiv\frac{1}{\lambda_3}\left(\frac{2\pi}{\Phi_0}\right)\frac{\sqrt{2}(L_3-M_{15}+M_{35})}{L_1(L_3+M_{35})-2M_{15}^2}\Phi_c^i$, is the ``frozen'' value of the fastest degree of freedom of qubit $i$. After performing the final rotation to align the slow degree of freedom with one of the coordinate directions, the pertinent terms of the interaction potential are given by
\begin{equation}
U^{AB}=\alpha q_1^A q_1^B+\beta( d_3^B\sin\theta^A q_1^A -d_3^A \sin\theta^B  q_1^B).\label{linear}
\end{equation}
As one can see, the linear terms in Eq. (\ref{linear}) are due to the ``frozen'' values of the fastest degree of freedom of each qubits. Moreover, the linear term of the slow coordinate of qubit $i$ arises due to the frozen coordinate of qubit $j$ and vice-versa. The interaction matrix elements due to these terms are exactly that of the one-qubit bias term, leading to a Hamiltonian form:   $$\frac{1}{2}B'(\Phi_c^B)b(\Phi_c^A)\hat{\sigma}_z^A\otimes\hat{\mathds{1}}^B-\frac{1}{2}B'(\Phi_c^A)b(\Phi_c^B)\hat{\mathds{1}}^A\otimes\hat{\sigma}_z^B$$

Finally, the term $q_1^A q_1^B$ of Eq. (\ref{linear}) only connects states of different parity in both qubits. Thus, this term leads the form:
$$J(\Phi_c^A,\Phi_c^B)\hat{\sigma}_z^A\otimes\hat{\sigma}_z^B$$

\section*{References}
\bibliographystyle{unsrt}
\bibliography{DC_paper_resub}
\end{document}